\documentclass[twocolumn]{aastex63}
\usepackage{ulem}
\usepackage{amsmath}
\graphicspath{{./fig/}}

\newcommand{\SigdotPE}{\dot{\Sigma}_{\rm PEW}}
\newcommand{\SigdotDW}{\dot{\Sigma}_{\rm MDW}}
\newcommand{\sub}[1]{_{\rm #1}}

\newcommand{\cwo}{C\sub{w,0}}
\newcommand{\cwe}{C\sub{w,e}}
\newcommand{\arp}{\overline{\alpha_{r\phi}}}
\newcommand{\apz}{\overline{\alpha_{\phi z}}}
\defcitealias{2020ApJ...892..124O}{Paper~I}

\shorttitle{Rapid disk clearing yields low-mass super-Earth atmospheres}
\shortauthors{Ogihara, Kunitomo and Hori}

\begin{document}

\title{Unified simulations of planetary formation and atmospheric evolution II:
Rapid disk clearing by photoevaporation yields low-mass super-Earth atmospheres}

\correspondingauthor{Masahiro Ogihara}
\email{masahiro.ogihara@nao.ac.jp}

\author[0000-0002-8300-7990]{Masahiro Ogihara}
\affiliation{National Astronomical Observatory of Japan,
2-21-1, Osawa, Mitaka,
181-8588 Tokyo, Japan}

\author[0000-0002-1932-3358]{Masanobu Kunitomo}
\affiliation{Kurume University,
67 Asahimachi, kurume,
830-0011 Fukuoka, Japan}

\author[0000-0003-4676-0251]{Yasunori Hori}
\affiliation{National Astronomical Observatory of Japan,
2-21-1, Osawa, Mitaka,
181-8588 Tokyo, Japan}
\affiliation{Astrobiology Center,
2-21-1, Osawa, Mitaka,
181-8588 Tokyo, Japan}

\begin{abstract}

Super-Earths possess low-mass H$_2$/He atmospheres (typically less than 10\% by mass). However, the origins of super-Earth atmospheres have not yet been ascertained.
We investigate the role of rapid disk clearing by photoevaporation during the formation of super-Earths and their atmospheres.
We perform unified simulations of super-Earth formation and atmospheric evolution in evolving disks that consider both photoevaporative winds and magnetically driven disk winds. For the growth mode of planetary cores, we consider two cases in which planetary embryos grow with and without pebble accretion.
Our main findings are summarized as follows. 
(i) The time span of atmospheric accretion is shortened by rapid disk dissipation due to photoevaporation, which prevents super-Earth cores from accreting massive atmospheres. 
(ii) Even if planetary cores grow rapidly by embryo accretion in the case without pebble accretion, the onset of runaway gas accretion is delayed because the isolation mass for embryo accretion is small. Together with rapid disk clearing, the accretion of massive atmospheres can be avoided.
(iii) After rapid disk clearing, a number of high-eccentricity embryos can remain in outer orbits. Thereafter, such embryos may collide with the super-Earths, leading to efficient impact erosion of accreted atmospheres.
We, therefore, find that super-Earths with low-mass H$_2$/He atmospheres are naturally produced by \textit{N}-body simulations that consider realistic disk evolution.
\end{abstract}


\section{Introduction} \label{sec:intro}

Recent observations, including those made by {\it Kepler} \citep{2010Sci...327..977B}, have found a large amount of low-mass planets around solar-type stars. Such planets are in the mass range between about 1 and $50 \,M_\oplus$, with typical core masses of about $3--10 \,M_\oplus$ \citep[e.g.,][]{2019ApJ...878...36L}, which are called super-Earths and/or sub-Neptunes. Hereafter we refer to these planets as super-Earths.
Several important orbital properties of super-Earths, including their period--ratio distributions, have been revealed \citep[e.g.,][]{2011ApJS..197....8L, 2014ApJ...790..146F}, which have been successfully reproduced by recent \textit{N}-body simulations of planet formation \citep[e.g.,][]{2017MNRAS.470.1750I, 2018A&A...615A..63O, 2019A&A...627A..83L}.
In this context, more information regarding the mass of H$_2$/He atmospheres of super-Earths is slowly being obtained. Although it is not currently possible to firmly constrain the amount of H$_2$/He atmospheres by transmission spectroscopy of transiting super-Earths \citep[e.g.][]{2013Sci...342.1473D}, there are several lines of evidence that most super-Earths do not possess massive H$_2$/He atmospheres. Combining the mass--radius relation with interior modeling, it is possible to estimate that most super-Earths possess low-mass atmospheres, typically less than about 10\% by mass \citep[][]{2014ApJ...792....1L}. In addition, the occurrence rate of super-Earths around solar-type stars is estimated to be about 30--50\% \citep[e.g.,][]{2013ApJ...766...81F, 2013PNAS..11019273P, 2018ApJ...860..101Z}, while those of hot-Jupiters and warm-Jupiters are expected to be less than 1\% \citep[][]{2003A&A...407..369U, 2010Sci...330..653H, 2011arXiv1109.2497M}, which indicates that super-Earths can easily avoid the runaway gas accretion process during the evolution of protoplanetary disks.

The formation of super-Earths with low-mass atmospheres is an open issue in planet-formation theory. One example of an unsolved problem is to explain why the cores of super-Earths, which are thought to undergo runaway gas accretion within the lifetime of the protoplanetary disk \citep{2012ApJ...753...66I, 2014ApJ...786...21P}, did not undergo runaway gas accretion. Several solutions have been proposed in the literature: high envelope opacity or polluted envelopes \citep[e.g.,][]{2014ApJ...797...95L, 2020A&A...634A..15B}, rapid recycling of the accreting gas \citep[e.g.,][]{2015MNRAS.447.3512O, 2017MNRAS.471.4662C, 2017A&A...606A.146L, 2018MNRAS.479..635K, 2019A&A...623A.179K}, limited disk accretion \citep[e.g.,][]{2018ApJ...867..127O, 2019MNRAS.487..681G}, atmospheric loss during disk dissipation \citep[e.g.,][]{2012ApJ...753...66I, 2020ApJ...889...77H}, and final assembly during disk dissipation \citep[e.g.,][]{2014ApJ...797...95L}.

To address this issue, we performed unified simulations of super-Earth formation and atmospheric evolution in \citet[][hereafter Paper~I]{2020ApJ...892..124O}, and found the following conclusions: (i) atmospheric heating by pebble accretion \citep{2014A&A...572A..35L}, which inhibits the inflow of disk gas, ceases before disk gas dispersal, (ii) atmospheric mass loss through giant impacts does not occur frequently, and (iii) massive atmospheres remain even after long-term photoevaporation of the atmosphere \citep[see also][]{2017ApJ...847...29O}. Therefore, it is suggested that other mechanisms are required to explain the formation of super-Earths with low-mass atmospheres.

In this paper, we consider a new disk evolution model that takes into account mass loss due to photoevaporation. In \citetalias{2020ApJ...892..124O}, we used a viscously evolving disk model in which mass loss due to magnetically driven disk winds (MDWs) and mass accretion due to disk winds (also known as wind-driven accretion) are considered \citep{2016A&A...596A..74S}.
In that study, mass loss due to photoevaporation was not included. This simplification may not be appropriate however, as disk gas is thought to be lost due to high-energy irradiation \citep[e.g.,][]{2001MNRAS.328..485C, Gorti+09, Tanaka+13, 2018ApJ...857...57N}. For example, \citet[][]{2020MNRAS.492.3849K} investigated the long-term evolution of protoplanetary disks by considering mass loss due to both magnetically driven winds and photoevaporative winds (PEWs). They showed that the disks can be rapidly depleted during the disk-dissipation phase, as also suggested by observations \citep[e.g.,][]{Haisch+01}. 

In this paper, we examine the effects of rapid disk clearing due to photoevaporation on the formation and evolution of super-Earth atmospheres. We perform \textit{N}-body simulations of super-Earth formation around a solar-mass star by adopting a new disk evolution model. In \citetalias{2020ApJ...892..124O}, we assumed that planetary cores grew mainly by pebble accretion. In this work, we also examine the case in which cores grow without pebble accretion.

The remainder of this paper is organized as follows. In Section~\ref{sec:model}, we outline our simulation model. In Section~\ref{sec:results}, we present the results of the simulations for disks that undergo, and do not undergo, mass loss due to photoevaporation. In Section~\ref{sec:discussion}, we discuss how super-Earths with low-mass atmospheres can form. Finally, in Section~\ref{sec:conc}, we summarize our conclusions.

\section{Model} \label{sec:model}

\subsection{Disk evolution}

We simulated the evolution of protoplanetary disks, including the effects of MDWs, PEWs, viscous accretion, and wind-driven accretion. As in \citetalias{2020ApJ...892..124O}, we numerically solved the 1D diffusion equation following the method of \citet{2016A&A...596A..74S}, in which the PEW effect was newly included here based on \citet{2020MNRAS.492.3849K}. We have briefly summarized our model in the following (see these papers for full details).

The equation for the density evolution is given by
\begin{align}
\frac{\partial \Sigma_{\rm g}}{\partial t} =& \frac{1}{r}\frac{\partial}{\partial r}
\left[\frac{2}{r\Omega}\left\{\frac{\partial}{\partial r}(r^2 \Sigma\sub{g}
  \overline{\alpha_{r\phi}}c_{\rm s}^2) + r^2 \overline{\alpha_{\phi z}}
  \frac{\Sigma_{\rm g}H\Omega^2}{2\sqrt{\pi}} \right\}\right]\nonumber\\
&- C_{\rm w} \frac{ \Sigma_{\rm g}\Omega }{2\sqrt{\pi}} - \SigdotPE
  \label{eq:sgmevl}
\end{align}
where $\Sigma_{\rm g}$ is the gas surface density, $\Omega$ is the angular velocity, $\rho$ is the density, and $c\sub{s}$ is the sound speed. In the simulations, we used cylindrical coordinates $(r, \phi, z)$, and we assumed $\Omega=\Omega\sub{K}$, where $\Omega\sub{K}$ is the Keplerian frequency.
The disk temperature was determined by stellar irradiation and viscous heating. For the latter, we considered energetics and adopt eqs.\,(19) and (23) from \citet[][]{2016A&A...596A..74S}. The terms $\arp$ and $\apz$ represent viscous accretion and wind-driven accretion, respectively. We also assumed that $\apz$ increased with decreasing $\Sigma\sub{g}$ \citep[see eq.\,(30) of ][]{2016A&A...596A..74S}.

The third term on the right-hand side of Eq.\,(\ref{eq:sgmevl}) represents the mass-loss rate via MDWs. The non-dimensional factor, $C_{\rm w}$, is given by
\begin{equation}\label{eq:cw}
C\sub{w}=\min \left( C\sub{w,0}, \cwe \right)\,,
\end{equation}
where $\cwo$ is a constant value inferred from local shearing box MHD simulations \citep{Suzuki+Inutsuka09,Suzuki+10}, and $C\sub{w}$ is also limited by energetics of accreting disks through $\cwe$. In this article, we adopted the value of $\cwe$ given in eq.\,(18) of \citet[][]{2016A&A...596A..74S}, which corresponds to  energetic winds (called a 'strong MDW' by \citet[][]{2016A&A...596A..74S})\footnote{The same prescription was used in \citetalias{2020ApJ...892..124O}.
}.

The last term on the right-hand side of Eq.\,(\ref{eq:sgmevl}) represents the mass-loss rate by PEWs, which in turn are driven by X-rays and extreme-UV (EUV) photons from the central star. The PEW rate, $\SigdotPE$, has been taken from the studies of \cite{Alexander+06a, Alexander+Armitage07,Owen+12}, as described by \citet{2020MNRAS.492.3849K}. We assumed typical values of the X-ray luminosity and EUV photon flux for solar-mass stars, which are $10^{30}\,\rm erg\,s^{-1}$ and $10^{41}\,\rm s^{-1}$, respectively.

In this article, the parameters of the disk evolution were chosen as $\arp = 8\times10^{-3}$ and $\cwo = 2\times10^{-5}$, unless otherwise noted. The stellar mass and bolometric luminosity were taken as $1\,{M_\odot}$ and $1\,{L_\odot}$, respectively. We also adopted $\Sigma\sub{g}\propto r^{-3/2}$ as an initial condition, and the initial disk mass was $0.118\,{\rm M_\odot}$. Please note that in this article, we assumed that planetary formation starts after the disk has evolved for 0.1\,Myr.

\begin{figure}[t!]
\epsscale{1.0}
\plotone{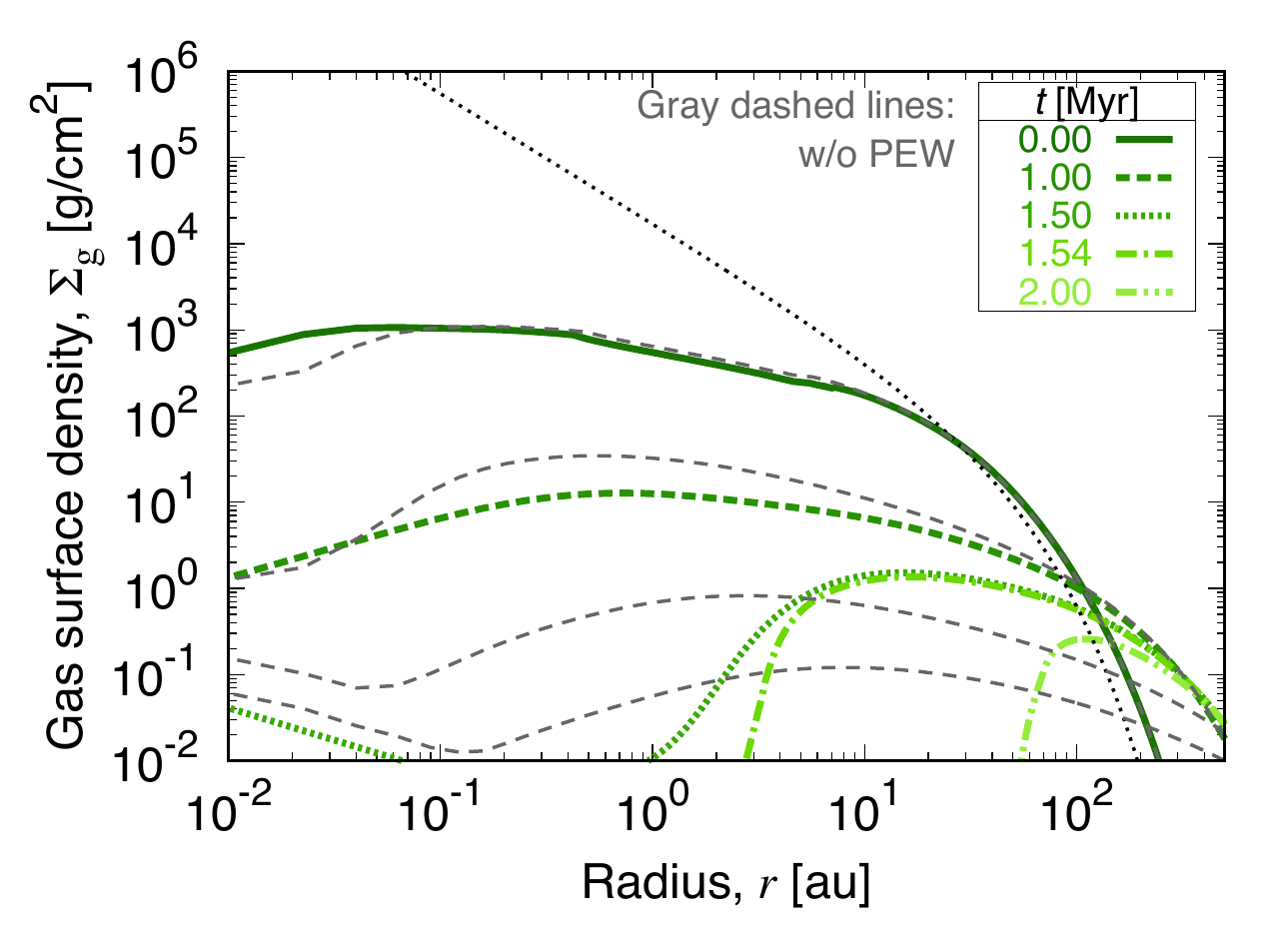}
\plotone{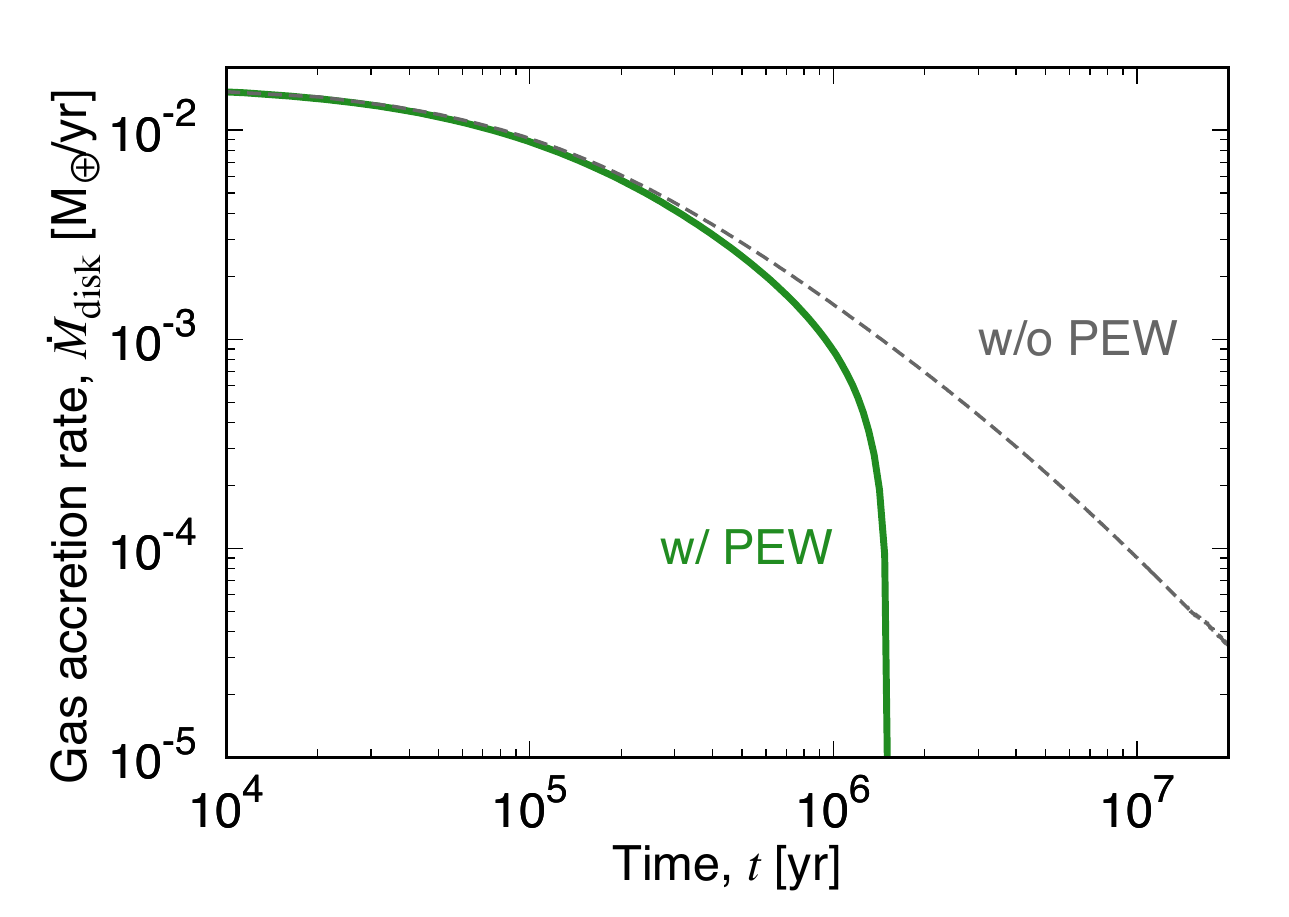}
\caption{
\textit{Top panel.} Temporal evolution of $\Sigma\sub{g}$. The thick green lines show the evolution with PEWs at $t=0, 1, 1.5, 1.54$ and 2\,Myr (from top to bottom), whereas the thin gray lines show the evolution without PEWs at $t=0, 1, 5$ and 10\,Myr (from top to bottom). Both coincide at $t=0$. The thin black dotted line shows the initial conditions of the disk evolution. We note that $t=0$ in this article corresponds to 0.1\,Myr after the disk evolution started (see text). \textit{Bottom.} Temporal evolution of the gas accretion rate , $\dot{M}_{\rm disk}$, at 0.3\,au in the cases with and without PEWs (solid green and dashed gray lines, respectively), which is discussed in more detail in Section~\ref{sec:discussion1}.
}
\label{fig:sgm_evol}
\end{figure}

Figure\,\ref{fig:sgm_evol} shows the evolution of the $\Sigma\sub{g}$ profile. Here, wind-driven accretion and $\SigdotDW$ played a dominant role in controlling the behavior of the disk gas, unless the accretion rate significantly decreased. There was no difference between the $t-\Sigma_{\rm g}$ profile with and without PEWs until $\simeq1\,$Myr \footnote{We used a time-explicit method to simulate the evolution with PEWs, as done by \citet{2020MNRAS.492.3849K}, whereas a time-implicit method was used for the case without PEWs, following \citetalias{2020ApJ...892..124O}. The different numerical schemes resulted in only small differences in $\Sigma_{\rm g}$ (at most a factor of three) in the early ($\lesssim 1\,$Myr) phase \citep[see also the discussion provided by ][]{2020MNRAS.492.3849K}.}.
At 1.54\,Myr, the X-ray driven PEWs opened a gap around 1\,au, and then the inner disk rapidly cleared out over the viscous timescale of the gap. The outer disk was directly irradiated, and completely dispersed within 2.5\,Myr. We refer to figure\,2 of \citet{2020MNRAS.492.3849K} for a case with a different set of parameters ($\arp = 8\times10^{-5}$ and $\cwo=1\times10^{-5}$), the values of which were used in the simulations shown in Appendix~\ref{sec:app}.

\subsection{Planetary formation}
The growth of planetary cores may proceed via pebble accretion and/or embryo accretion. In this article, we have examined both cases.

\subsubsection{Pebble accretion}

One theory is that planets grow by accreting pebbles. Here, we briefly summarize prescriptions for pebble accretion (see also section~2.1.1 of \citetalias{2020ApJ...892..124O} for more details). In the simulations that considered pebble accretion, we started with embryos of $M_{\rm ini} = 0.01 \,M_\oplus$. These masses were larger than the transition mass from the Bondi regime to the Hill regime \citep[][]{2012A&A...544A..32L}. Thus, we basically considered the Hill regime only. The pebble accretion rate is
\begin{eqnarray}\label{eq:mdot_pb}
\dot{M}_{\rm acc,pb} = \left\{ \begin{array}{ll}
    2 r_{\rm eff} v_{\rm acc} \Sigma_{\rm pb} & ({\rm for ~2D}) \\
    \pi r_{\rm eff}^2 v_{\rm acc} \rho_{\rm pb} & ({\rm for ~3D}),
  \end{array} \right.
\end{eqnarray}
where $\Sigma_{\rm pb}$ and $\rho_{\rm pb}$ are the pebble surface density and the pebble density on the midplane, respectively. In the 2D (or 3D) accretion mode, the pebble scale height $H_{\rm pb}$ is smaller (or larger) than the effective cross section $r_{\rm eff}$. The accretion velocity for the Hill regime is given by $v_{\rm acc} \simeq r_{\rm eff} \Omega_{\rm K}$. The effective cross section of pebbles for the Hill regime is
\begin{equation}
r_{\rm eff} = \left(\frac{\tau_{\rm s}}{0.1}\right)^{1/3} R_{\rm H},
\end{equation}
where $\tau_{\rm s}$ and $R_{\rm H}$ indicate the Stokes number and the Hill radius, respectively. When the stopping time of pebbles was longer than the inverse of the Keplerian frequency, a reduction of $r_{\rm eff}$ was considered \citep[e.g.,][]{2010A&A...520A..43O,2012ApJ...747..115O}.
The pebble accretion efficiency can change when eccentricities and inclinations of planets are excited \citep{2018A&A...615A.138L,2018A&A...615A.178O}, which is not considered here for simplicity.

For the pebble surface density, we adopted the same model as \citetalias{2020ApJ...892..124O}, such that
\begin{equation}
\Sigma_{\rm pb} = \frac{\dot{M}_{\rm pb}}{2 \pi r v_r}.
\end{equation}
The radial drift velocity $v_{\rm r}$ is given following \citep{1977MNRAS.180...57W,1986Icar...67..375N}
\begin{equation}
v_r = -\frac{2\tau_{\rm s}}{\tau_{\rm s}^2 +1} \eta v_{\rm K},
\end{equation}
where $\eta$ represents the non-dimensional pressure gradient. As the value of the pebble mass flux is poorly constrained, the pebble mass flux, $\dot{M}_{\rm pb}$, is treated as a parameter \citep[e.g.,][]{2019A&A...627A..83L, 2019arXiv190208772I, 2019A&A...623A..88B}. We considered 1-mm-size silicate pebbles, as in previous studies \citep[e.g.,][]{2015Icar..258..418M, 2019arXiv190208772I}. See also discussions in section 2.2 of \citet{2019A&A...632A...7L}.

The growth timescale of 2D pebble accretion is roughly given by
\begin{eqnarray}\label{eq:tacc_pb}
t_{\rm acc,pb} = \frac{M}{\dot{M}_{\rm acc,pb}} 
&\simeq& 1 \times 10^5\,{\rm yr}
\left(\frac{\tau_{\rm s}^{1/3}}{\tau_{\rm s}^2 + 1} \right)
\left( \frac{M}{M_\oplus}\right)^{1/3}\nonumber \\
&&\times \left(\frac{\dot{M}_{\rm pb}}{10^{-4} M_\oplus {\rm \,yr^{-1}}} \right)^{-1} 
\left( \frac{\eta}{10^{-3}}\right).
\end{eqnarray}

The accretion of pebbles is halted when the planet reaches the pebble isolation mass \citep[e.g.,][]{2012A&A...546A..18M,2018A&A...612A..30B}
\begin{eqnarray}\label{eq:miso_pb}
M_{\rm iso,pb} &=& 25 \,M_\oplus \left( \frac{H/r}{0.05}\right)^3 \left\{ 0.34 \left( \frac{-3}{\log{10}(\arp)}\right)^4 + 0.66 \right\} \nonumber \\
&& \times \left( 1 - \frac{ \frac{\partial \ln P}{\partial \ln r}+ 2.5}{6}\right).
\end{eqnarray}
Although it is usually assumed that wind-driven accretion may not contribute to the pebble isolation mass \citep[e.g.][]{2019A&A...622A.202J}, it could be possible that the pebble isolation mass becomes larger due to $\apz$. Note that the increase in the pebble isolation mass does not change our conclusions in this paper. The effect of wind-driven accretion on the pebble isolation mass should be investigated by hydrodynamical simulations.

\subsubsection{Embryo accretion}

In half of the simulations performed in this work, we considered core growth by embryo collisions, rather than pebble accretion. We started these simulations with an embryo mass of $M_{\rm ini} = 0.1 \,M_\oplus$, unless otherwise stated.
The growth rate of a core in the dispersion-dominated regime in a swarm of embryos with a local surface density ($\Sigma_{\rm d}$) and velocity dispersion ($v$) is
\begin{eqnarray}
\dot{M}_{\rm acc,em} = \Sigma_{\rm d} \Omega_{\rm K} \pi R^2 \left( 1 + \frac{v_{\rm esc}^2}{v^2}\right),
\end{eqnarray}
where $R$ and $v_{\rm esc} (=\sqrt{2GM/R})$ are the physical radius and the surface escape velocity of the core, respectively.

Then, the core growth timescale by embryo accretion is
\begin{eqnarray}\label{eq:tacc_pl}
t_{\rm acc,em} = \frac{M}{\dot{M}_{\rm acc,em}} 
&\simeq& 6 \times 10^6 {\rm \,yr}
\left(\frac{\Sigma_{\rm d}}{77{\rm \,g \,cm^{-2}}} \right)^{-1}
\left(\frac{a}{1{\rm \,au}} \right)^{3/2}\nonumber \\
&& \times \left(\frac{M}{M_\oplus} \right)^{1/3}
\left(1 + \frac{v_{\rm esc}^2}{v^2} \right)^{-1},
\end{eqnarray}
where we assumed that the mean density of an embryo was $ 3\, \rm g\,cm^{-3}$.

Core growth by embryo accretion is halted when all of the embryos in the vicinity of the core have accreted. This isolation mass is
\begin{eqnarray}\label{eq:miso_pl}
M_{\rm iso,em} &=& 2 \pi a \Delta a \Sigma_{\rm d},\\
&\simeq& 2.4 \,M_\oplus
\left(\frac{C}{10}\right)^{3/2}
\left(\frac{\Sigma_{\rm d}}{77\,{\rm g \,cm^{-2}}}\right)^{3/2}
\left(\frac{a}{1 {\rm \,au}}\right)^{3}\nonumber\\
&&\times \left(\frac{M_*}{M_\odot}\right)^{-1/2},\label{eq:miso_pl2}
\end{eqnarray}
where $\Delta a = C R_{\rm H}$, and $C \simeq 10$ for oligarchic growth. 
Note that the isolation mass is given under the assumption that planets stay their orbits. As we show in the following sections, migration is a minor effect in our simulations.

\subsubsection{Collisions}\label{sec:collision}
The orbital evolutions of all bodies in the \textit{N}-body simulations were tracked, and the mutual gravitational interactions between them were calculated. In standard models, perfect merging is assumed. In some simulations, we also considered hit-and-run collisions.
According to \citet{2012ApJ...744..137G}, the critical impact velocity for perfect accretion between planets with masses of $M_1$ and $M_2$ is
\begin{eqnarray}
\frac{v_{\rm crit}}{v_{\rm esc,2b}} &=& 2.43 \left( \frac{M_1 - M_2}{M_1 + M_2} \right)^2 (1- b)^{5/2} \nonumber \\
&& - 0.0408 \left( \frac{M_1 - M_2}{M_1 + M_2} \right)^2 + 1.86 (1- b)^{5/2} \nonumber \\
&& + 1.08,
\end{eqnarray}
where $v_{\rm esc,2b}$ is the two-body escape velocity. The impact angle is expressed by the impact parameter $b (= \sin \theta)$, where $b = 0$ means a head-on collision. When the impact velocity, $v_{\rm imp}$, was smaller than the critical velocity, the collisions were treated as perfect mergers. Otherwise, the planets 'bounced' with a tangential velocity of $v_{\rm t}^\prime = \max(v_{\rm imp,t}, v_{\rm esc,2b})$, where $v_{\rm imp,t}$ is the tangential component of the impact velocity.

\subsubsection{Disk--planet interactions}

Mars-mass or larger planets undergo damping of their semi-major axes and eccentricities ($e$) due to disk--planet interactions. In this work, we used the formula for type-I migration developed by \citet{2011MNRAS.410..293P}. Planets undergo rapid inward migration in a power-law disk, e.g. like a minimum-mass solar nebula. However, as the surface density slope can be flat or even positive in our disk model, the desaturated co-rotation torque could yield a positive torque, thereby leading to suppressed inward migration \citep[][]{2018A&A...615A..63O}. 
We also considered a reduction of dampings for planets with high values of $e$ and $i$ \citep[][]{2008A&A...482..677C}.
Type-II migration for larger planets was also considered. According to \citet{2018ApJ...861..140K}, the type-II migration timescale can be expressed as $t_{a,{\rm II}} \simeq t_{a,{\rm I}} (\Sigma_{\rm g}/\Sigma_{\rm min})$, where $\Sigma_{\rm min}$ is the gas surface density at the bottom of the gap, which is expressed by
\begin{eqnarray}\label{eq:sigma_min}
\frac{\Sigma_{\rm g}}{\Sigma_{\rm min}} = 1 + 0.04 \left( \frac{M}{M_*} \right)^2 \left( \frac{H}{r} \right)^{-5} \arp^{-1}.
\end{eqnarray}

\subsection{Atmospheric evolution}

\subsubsection{Atmospheric accretion}\label{sec:atm_acc}

For a planetary core to start accreting a massive H$_2$/He atmosphere from the gas disk, the cores should exceed the critical core mass. The critical core mass increases as the atmosphere becomes heated due to the accretion of pebbles. In \citetalias{2020ApJ...892..124O}, we performed a series of 1D structure calculations with accretion heating, and derived the following formula
\begin{eqnarray}\label{eq:mcrit}
M_{\rm crit} = 13 \,M_\oplus \left(\frac{\dot{M}_{\rm acc,pb}}{10^{-6} \,M_\oplus \, {\rm yr}^{-1}}\right)^{0.23}.
\end{eqnarray}
A detailed explanation is given in the appendix of \citetalias{2020ApJ...892..124O}.

The actual atmospheric accretion rate onto a core that exceeds the critical core mass is given by
\begin{eqnarray}\label{eq:mdot_min}
\dot{M}_{\rm atm} = \min(\dot{M}_{\rm KH}, \dot{M}_{\rm hydro}, \dot{M}_{\rm disk}).
\end{eqnarray}
The first term on the right-hand side is determined by Kelvin--Helmholtz contraction \citep{2010ApJ...714.1343H}, such that 
\begin{equation}
\dot{M}_{\rm KH} = 10^{-8} \,M_\oplus \,{\rm yr}^{-1} \left(\frac{M_{\rm core}}{M_\oplus}\right)^{3.5}.
\end{equation}
The second term represents the gas-capture rate derived from hydrodynamic simulations \citep{2016ApJ...823...48T}
\begin{eqnarray}
\dot{M}_{\rm hydro} &=& 0.29 \left(\frac{H}{r}\right)^{-2} \left(\frac{M}{M_*}\right)^{4/3} r^2 \Omega_{\rm K} \Sigma_{\rm min},
\end{eqnarray}
where $\Sigma_{\rm min}$ is expressed in Eq.~(\ref{eq:sigma_min}).
The third term represents the supply limit of the global disk accretion, which is calculated by
\begin{equation}\label{eq:mdot_disk}
\dot{M}_{\rm disk} = \max(\dot{M}_{\rm visc}, \dot{M}_{\rm wind}),
\end{equation}
where $\dot{M}_{\rm visc}$ and $\dot{M}_{\rm wind}$ are viscous accretion and wind-driven accretion, respectively. See section~2.1.4 of \citetalias{2020ApJ...892..124O} for more details.
Note that $\dot{M}_{\rm KH}$ becomes smaller for smaller values of $M_{\rm core}$. The Kelvin--Helmholtz timescale is longer than the typical disk lifetime considered in this work ($2 {\rm \,Myr}$) for cores with $M_{\rm core} \lesssim 3 \,M_\oplus$. Therefore, from here on we consider the mass $M=\min(\dot{M}_{\rm crit}, 3 \,M_\oplus)$ as the critical core mass.

\subsubsection{Atmospheric losses}

An accreted atmosphere can be blown off during an impact with another planet \citep[e.g.,][]{2003Icar..164..149G,2015Icar..247...81S}. As in \citetalias{2020ApJ...892..124O}, we used a scaling law obtained from giant-impact simulations \citep{2014LPI....45.2869S} .
The atmospheric loss fraction ($L$) of a planet with $M_{\rm core}=M_1$ is approximately given by
\begin{eqnarray}
 L = \left\{ \begin{array}{ll}
    1 & (10^{7.76} < Q_{\rm S} ) \\
    0.562 \log_{10}{Q_{\rm S}} - 3.37& (10^{6.35} < Q_{\rm S} \leq 10^{7.76})\\
    0.0850 \log_{10}{Q_{\rm S}} - 0.340& (Q_{\rm S} \leq 10^{6.35}),
  \end{array} \right.
\end{eqnarray}
where $Q_{\rm S}$ is the specific impact energy of a collision with a planet of $M_{\rm core}=M_2$, which is given by
\begin{eqnarray}\label{eq:Qs}
Q_{\rm S} = Q_{\rm R} \left( 1 + M_2/M_1\right) (1-b),\\
Q_{\rm R} = \frac{1}{2} \frac{M_1 M_2}{(M_1 + M_2)^2} v_{\rm imp}^2.
\end{eqnarray}
This scaling law can be derived for terrestrial planets with relatively thin atmospheres. Although the proportion of atmospheric erosion decreases for massive atmospheres, a similar scaling law was obtained in recent SPH simulations by \cite{2020arXiv200202977K}. In our simulations, when a planet underwent a hit-and-run collision, we assumed that same scaling law held.

\section{Results}\label{sec:results}

\subsection{Pebble accretion}\label{sec:pebble}

\subsubsection{In a disk with no PEW effects}

\begin{deluxetable}{llcl}
\tablecaption{List of models. Five runs were performed for each model. 
\label{tbl:list}}
\tabletypesize{\footnotesize}
\tablehead{
\colhead{Name} &  \colhead{PEW} &  \colhead{ $\dot{M}_{\rm pb} (M_\oplus/{\rm yr}$)} &  \colhead{Comments}
} 
\startdata
PB1 &  No  & $1.0 \times 10^{-4}$ & Without PEWs (same as in \citetalias{2020ApJ...892..124O}) \\
PB2 &  Yes & $1.0 \times 10^{-4}$ & Standard model\\
PB3 &  Yes & $2.0 \times 10^{-4}$ & High pebble flux\\
PB4 &  Yes & $1.0 \times 10^{-4}$ & Hit-and-run collisions\\
PB5 &  No  & $2.0 \times 10^{-4}$ & High pebble flux for PB1
\enddata
\end{deluxetable}

\begin{figure*}[ht!]
\epsscale{1.1}
\plotone{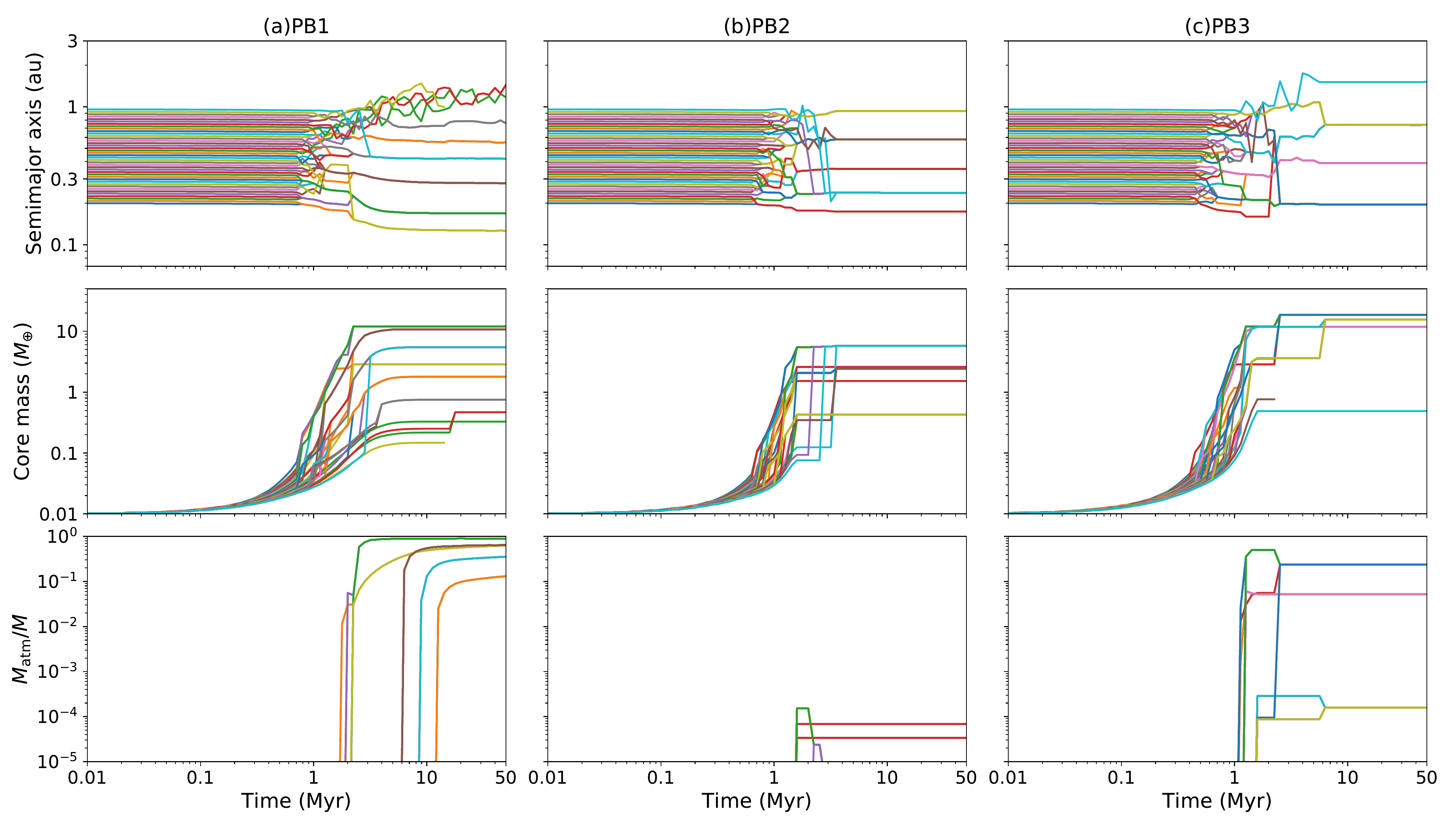}
\caption{Temporal evolution of the semi-major axes ({\it top}), core masses ({\it middle}) and atmospheric mass fractions ({\it bottom}).
}
\label{fig:t_a1}
\end{figure*}

First, let us review the results of the simulations that considered pebble accretion. The conditions of each model are listed in Table~\ref{tbl:list}.
Simulations started with embryos with $0.01 \,M_\oplus$ that were distributed between 0.2 and 1 au with orbital separations of 15 mutual Hill radii, which is comparable those used in \citetalias{2020ApJ...892..124O}.
Figure~\ref{fig:t_a1}(a) shows a typical outcome of our simulations in a disk without the effects of PEW (PB1), which is comparable to the simulations conducted by \citetalias{2020ApJ...892..124O}. To reduce the computational cost, the range of orbital distances of the initial planetary seeds was narrower in this paper (i.e., initial seeds are distributed between 0.1 and 2 au in \citetalias{2020ApJ...892..124O}). Nevertheless, the results are basically the same as found by \citetalias{2020ApJ...892..124O}.

A brief overview of the orbital evolution in this scenario is as follows. For more details, the reader is referred to the work of \citet{2018A&A...615A..63O} and \citetalias{2020ApJ...892..124O}. We found that type-I migration was significantly suppressed, which have been due to the surface density slope being almost flat in the inner region at $r < 1{\rm \,au}$ (see Figure~\ref{fig:sgm_evol}). In addition, the gas surface density is much lower than the minimum-mass solar nebula, which also contributes to the suppression of migration.
Planets with masses of $M\gtrsim 0.1\,M_\oplus$ underwent slow inward migration after $t \sim 1 \,{\rm Myr}$, and some planets were captured in mean-motion resonances (MMRs). In the simulation shown in Figure~\ref{fig:t_a1}(a), the planets formed a chain of MMRs (typically 5:4 MMRs) at $t \simeq 1.5 {\rm \,Myr}$. The resonant chain exhibited an orbital instability at $t \simeq 2.2 {\rm \,Myr}$, in which the chain had broken. We performed five runs for each model, where we changed the initial locations of embryos in each, and we found that only two pairs of MMRs remained at the end of all simulations. This fraction of resonant systems is consistent with the orbital properties of observed super-Earths \citep[][]{2018A&A...615A..63O}.

\begin{figure*}[ht!]
\epsscale{1.1}
\plotone{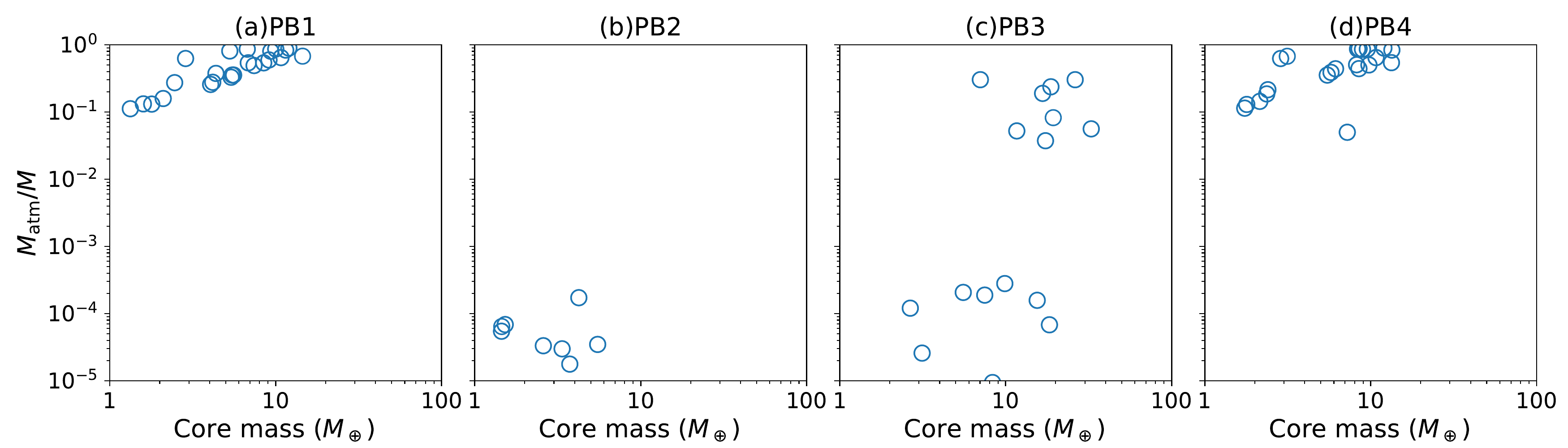}
\caption{Atmospheric mass fractions at the end of the simulations (t = 50 {\rm \,Myr}). Five simulations were performed for each model.
}
\label{fig:mcore_menv}
\end{figure*}

The evolution of atmospheric masses simulated here was the same as considered by \citetalias{2020ApJ...892..124O}. As the planetary cores grew via pebble accretion, the critical core mass was kept large $(\gtrsim 10 {\rm \,M_\oplus})$ due to the accretion heating of the atmospheres (Eq.~\ref{eq:mcrit}). Some planets reached the pebble isolation mass by $t \simeq 1.6 {\rm \,Myr}$, and after that, the critical core masses became smaller than the core masses (i.e., $M_{\rm core} \gtrsim 3 \,M_\oplus$ while $M_{\rm crit} \simeq 3 \,M_\oplus$), which led to the accretion of massive H$_2$/He atmospheres. This process was also confirmed in the recent work \citep{2020arXiv200604121B}. The late orbital instability at $t \simeq 2.2 {\rm \,Myr}$ mentioned above was likely triggered by planetary growth (i.e., atmospheric accretion, and core accretion by pebbles).
Figure~\ref{fig:mcore_menv} summarizes the final atmospheric mass fractions of the five runs of simulations for each model. In the simulations that considered disks without PEW (panel~a), the super-Earth cores accreted massive ($\simeq 10--90 {\rm \,wt\%}$) H$_2$/He atmospheres, which is inconsistent with the H$_2$/He atmospheric proportions observed for known super-Earths.

\subsubsection{In a disk with PEW effects}\label{sec:pebble_PEW}

Figure~\ref{fig:t_a1}(b) shows a typical outcome in a disk under the effects of PEWs (PB2). We found that the accretion of massive atmospheres could be avoided in this case. The time taken to reach the pebble isolation mass ($t \simeq 1.5 {\rm \,Myr}$) was comparable to that in PB1 simulations. However, the disk gas rapidly dissipated due to PEWs at $t \gtrsim 1.5 {\rm \,Myr}$ (see Figure~\ref{fig:sgm_evol}). Therefore, the planets ceased accreting atmospheres soon after they exceeded the critical core mass.
Figure~\ref{fig:mcore_menv}(b) summarizes the five runs of simulations for this model. We confirm that no planets accreted massive atmospheres. This is consistent with the estimated proportion of atmospheres of observed super-Earths ($\lesssim 10 {\rm wt\%}$). Note that the simulated planets possessed only very low-mass H$_2$/He atmospheres ($< 0.01\%$ mass fraction); however, the actual atmospheric fractions could be slightly higher because we did not consider atmospheric accretion onto cores that were smaller than the critical core mass.

Figure~\ref{fig:t_a1}(c) shows the results of simulations (PB3) in which the pebble flux increased by a factor of two from the simulation shown in Figure~\ref{fig:t_a1}(b) (i.e., PB2). In this case, the cores reached the pebble isolation mass at $t \simeq 1.0 {\rm \,Myr}$, and the time taken to reach the pebble isolation mass shortened by $\simeq 0.5 {\rm \,Myr}$. Thus, some planets that reached the pebble isolation mass accreted some gaseous material before the inner gas disk became depleted of gas at $t \gtrsim 1.5 {\rm \,Myr}$.
In Figure~\ref{fig:mcore_menv}(c), we see that several planets accreted atmospheres with more than 10wt\%. Note, however, that the proportions of accreted atmospheres were smaller than those found for PB1 without PEW effects, even though the cores were more massive. This is because the duration of atmospheric accretion was shortened by the rapid disk clearing. Thus, the mass loss due to PEWs played an important role in decreasing the amount of accreted atmospheric gas.
In Figure~\ref{fig:mcore_menv}(c), we can also see that smaller planets with $M_{\rm core} \lesssim 10 \,M_\oplus$ did not accrete thick atmospheres. These planets grew in wider orbits ($r\simeq 1{\rm \,au}$), where the pebble-accretion rate was smaller, and the time to reach the pebble isolation mass was almost comparable to the timing of disk depletion. 

We also investigated the effects of hit-and-run collisions on the amount of H$_2$/He atmospheres. When hit-and-run collisions were considered, the planets experienced a larger number of impact events to grow to the same size. Figure~\ref{fig:mcore_menv}(d) shows a summary of simulations performed without PEW effects, but in which hit-and-run collisions were considered (PB4). We found that planets in PB4 showed final atmospheric mass distributions similar to those found for the simulations in which perfect accretion between embryos was assumed (PB1). This is because, as shown in Figure~\ref{fig:t_a1}(a), the planets were in stable orbits after accreting massive atmospheres ($t > 3 {\rm \,Myr}$), and giant-impact events were rare during this phase. We also found that hit-and-run collisions were less frequent than collisional events of perfect merging. In fact, there was no clear difference in the total number of erosion events between PB1 and PB4.

\begin{figure}[ht!]
\epsscale{1.0}
\plotone{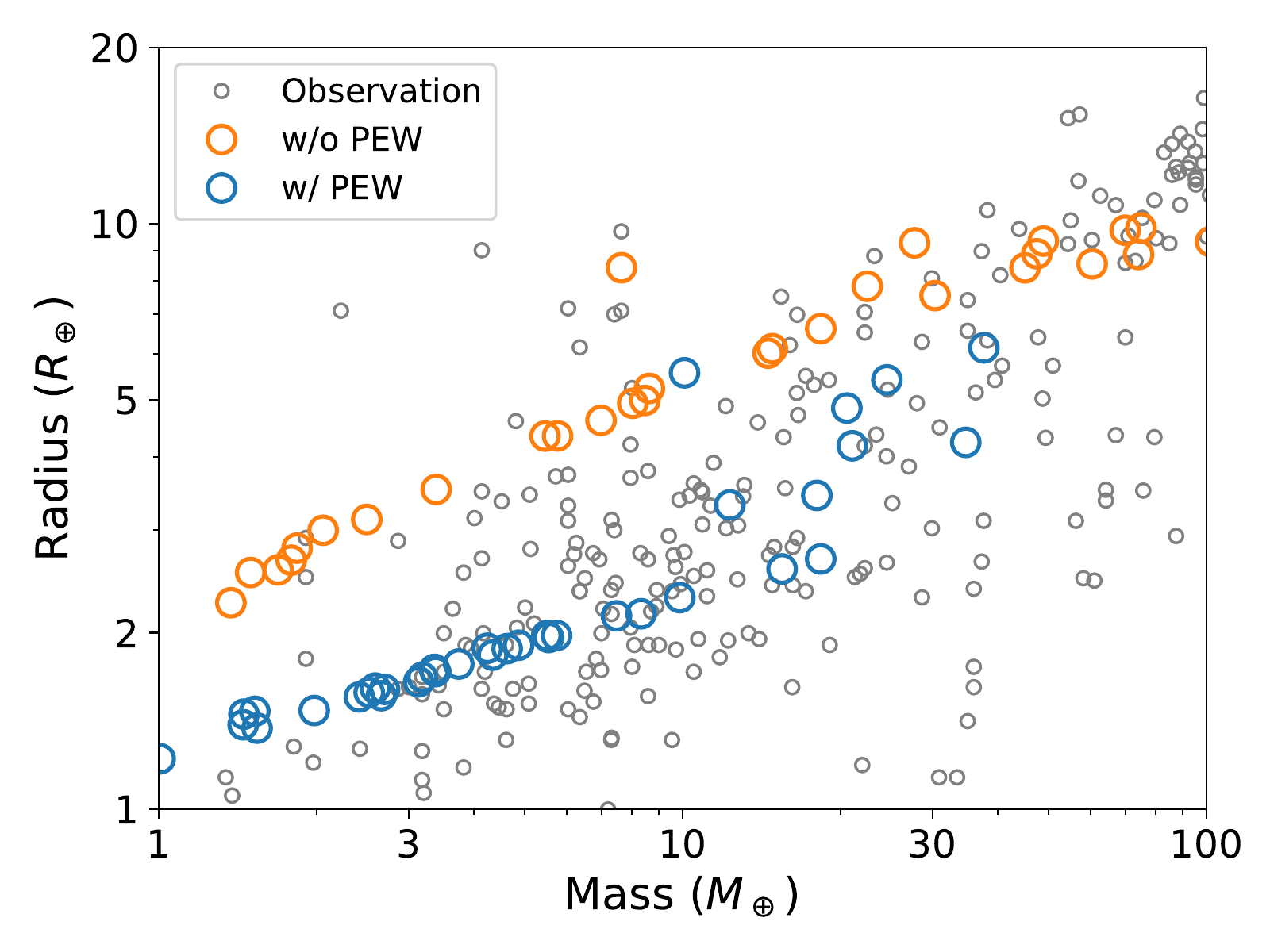}
\caption{Mass--radius diagram of the simulated planets, in which the small circles represent confirmed exoplanets. Blue circles show the results of simulations PB2 and PB3 (i.e., with PEW effects) at $t=50 {\rm \,Myr}$, and orange circles are the results from PB1 and PB5 (i.e., without PEW effects), where the pebble flux for PB5 was twice as large as that for PB1.}
\label{fig:m_r1}
\end{figure}

Next, we compared the mass--radius distributions of the simulated planets with those of observed planets. Figure~\ref{fig:m_r1} shows a mass--radius diagram, in which the small circles represent confirmed exoplanets, and the blue circles show the results of simulations PB2 and PB3 at the end of simulation ($t=50 {\rm \,Myr}$). To compare with PB3, we performed additional simulations (PB5) in which the pebble flux was increased by a factor of two from that used in PB1. The orange circles in Figure~\ref{fig:m_r1} thus show the results of simulations in disks without mass loss due to PEWs (PB1 and PB5). 
The radii of the simulated planets were calculated by interior models (see the method presented by \citetalias{2020ApJ...892..124O}). The planetary radii of the planets that formed in disks that did not undergo mass loss due to PEWs (PB1 and PB5) were much larger than those of observed super-Earths. On the other hand, we found that the radii of the formed planets for PB2 and PB3 were consistent with the typical radii of observed super-Earths. In the PB3 simulations, where the pebble flux was increased by a factor of two, planets with $M\simeq 10-40 \,M_\oplus$ possessed some amount of atmospheric gas; however, their radii were smaller than those found in the PB1 and PB5 simulations.

\subsection{Embryo accretion}\label{sec:planetesimal}

\begin{deluxetable}{llcl}
\tablecaption{List of models. Five runs were performed for each model. We adopted a power-law distribution for the initial solid distribution as $\Sigma_{\rm d} = \Sigma_{\rm d,ini} (r/1{\rm \,au})^{-3/2}$. \label{tbl:list2}}
\tabletypesize{\footnotesize}
\tablehead{
\colhead{Name} &  \colhead{PEW} &  \colhead{ $\Sigma_{\rm d,ini} ({\rm g/cm^2}$)} &  \colhead{Comments}
} 
\startdata
EM1 &  No   & $ 77   $ & Without PEWs\\
EM2 &  Yes  & $ 77   $ & Standard model\\
EM3 &  Yes  & $ 154 $ & Large solid amount\\
EM4 &  Yes  & $ 154 $ & $M_{\rm ini}=0.03\,M_\oplus$\\
EM5 &  No   & $  77  $ & $M_{\rm ini}=0.03\,M_\oplus$ for EM1
\enddata
\end{deluxetable}

\begin{figure*}[ht!]
\epsscale{1.1}
\plotone{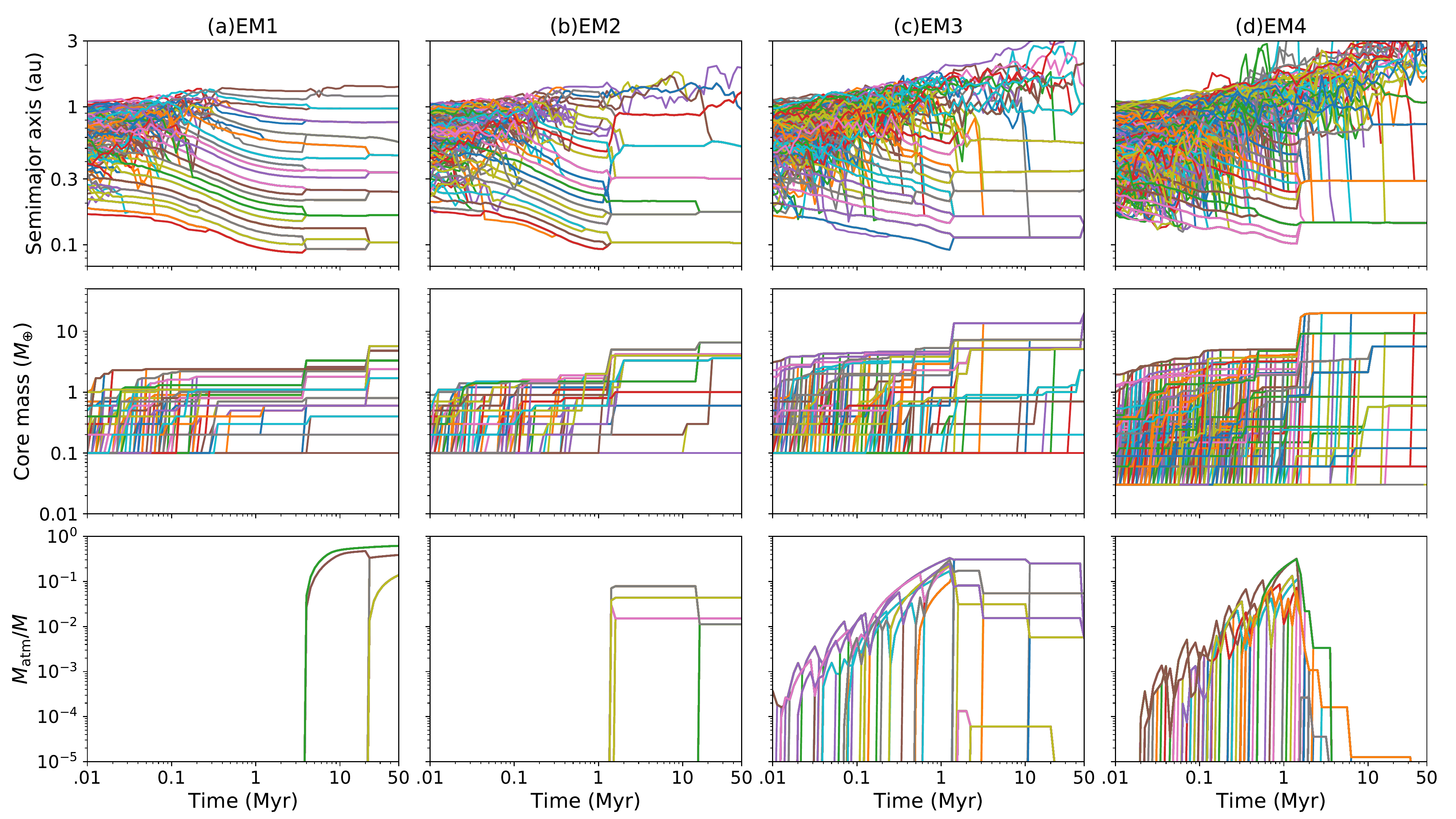}
\caption{Same as Figure~\ref{fig:t_a1}, but for models EM1--EM4 in which planets grew solely by planetesimal accretion.}
\label{fig:t_a2}
\end{figure*}

\begin{figure*}[ht!]
\epsscale{1.1}
\plotone{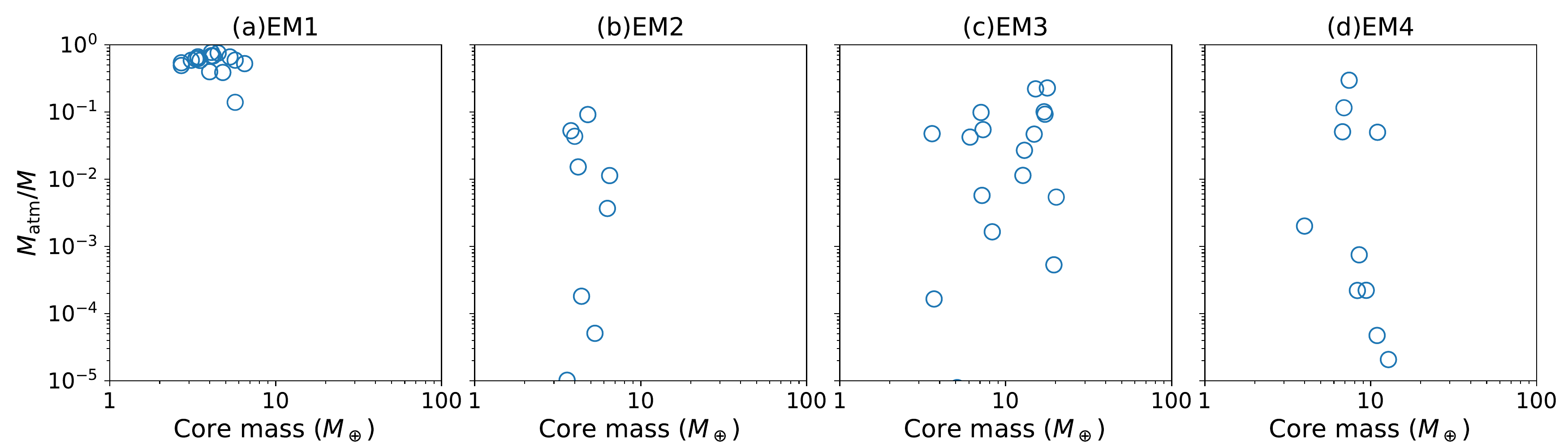}
\caption{Same as Figure~\ref{fig:mcore_menv}, but for models EM1--EM4.}
\label{fig:mcore_menv2}
\end{figure*}

\subsubsection{In a disk without PEW effects}

In this section, we present the results of simulations in which the planets grew solely by collisions with embryos. The properties of each model are listed in Table~\ref{tbl:list2}. Embryos with $0.1 \,M_\oplus$ (except for EM4 and EM5) were initially distributed in accordance with $\Sigma_{\rm d} \propto r^{-3/2}$. The total initial solid mass was assumed as $20 \,M_\oplus$ for the fiducial case of $\Sigma_{\rm d} = 77 {\rm \,g \,cm^{-2}} (r/1 {\rm \,au})^{-3/2}$.
Figure~\ref{fig:t_a2}(a) shows a typical outcome of our simulation for disks that did not experience mass loss due to PEWs (EM1). The big difference with the pebble-accretion model is that the cores here grew more rapidly, and reached $1 \,M_\oplus$ before $t=0.01--0.1 {\rm \,Myr}$. This is because the solid surface density, $\Sigma_{\rm d}$, was large at the beginning of the simulation. 
The orbital evolution of EM1 was found to be the same as that of \citet{2018A&A...615A..63O}, and we did not observe rapid type-I migration. The effects of inward migration were clearer than that seen in Figure~\ref{fig:t_a1}(a) (PB1) because Earth-mass cores formed in the early massive-disk phase at $t < 0.1 {\rm \,Myr}$. Nevertheless, the migration speed was orders of magnitude smaller than for classical power-law disks, such as a minimum-mass solar nebula \citep[e.g.,][]{2015A&A...578A..36O}. During the slow-migration phase, most planets were captured in close MMRs (e.g., 6:5 MMRs). As the disk gas slowly dissipated, the system experienced a late orbital instability after $t \simeq 3 {\rm \,Myr}$, which led to a breakup of the resonant chain. Note that the late orbital instability was usually triggered by planetary growth (i.e., atmospheric accretion or pebble accretion) in the simulations with pebble accretion (e.g., PB1, PB2). However, in this case, the late instability was triggered by disk dissipation. As seen in the pebble-accretion simulations, most planets were not in MMRs in the final state. This is consistent with the orbital properties of observed super-Earths.

The temporal evolution of the core masses, and the amount of accreted H$_2$/He atmospheres, are shown in Figure~\ref{fig:t_a2}(a). As the growth timescale was short, the cores reached the isolation mass ($\lesssim 2 \,M_\oplus$) at $t = 0.01--0.1 {\rm \,Myr}$. Note that the cores with isolation masses were not large enough to start runaway gas accretion, because $M_{\rm crit}$ was larger than $3 \,M_\oplus$. That is, the atmospheric accretion timescale is longer than the disk lifetime (Section~\ref{sec:atm_acc}).
They did not grow further because they keep separations between neighboring cores larger than $\simeq 6 \,R_{\rm H}$. During the giant-impact phase in the dissipating disk ($t > 3 {\rm \,Myr}$), the cores grew large enough to accrete massive atmospheres. After the cores exceeded the critical core mass, there existed some remnant disk gas, and hence disk accretion occurred (see Figure~\ref{fig:sgm_evol}).
In the final state, the planets possessed massive H$_2$/He atmospheres, which is the same as shown in Figure~\ref{fig:t_a1}(a). Note that although the gas surface density of the disks was already small ($\sim 1 {\rm \,g \,cm^{-2}}$) after the giant impacts, the planets were still able to accrete enough gas from the remnant disk to form atmospheres. For more details, see the discussion presented in Section~\ref{sec:discussion1}.
Figure~\ref{fig:mcore_menv2} summarize five runs of simulations for each model. We found that the super-Earth cores had thick ($\gtrsim 50$ wt\%) atmospheres, as seen in panel (a) for EM1.

\subsubsection{In a disk with PEW effects}

Figure~\ref{fig:t_a2}(b) shows typical outcomes of simulated disks that evolve with mass loss due to PEWs (EM2). As shown in Figure~\ref{fig:t_a2}(a) for EM1, although the cores grew rapidly, cores with masses equal to the isolation mass ($\lesssim 2 \,M_\oplus$) did not start runaway gas accretion before the disk dissipated. After the disk gas dissipated to some extent, giant impacts between cores occurred, and the cores started accreting atmospheres thereafter. Note that the giant-impact events occurred earlier ($t \simeq1.2{\rm \,Myr}$) than in EM1, because the disk dissipated earlier in this case (see Figure~\ref{fig:sgm_evol}). Soon after that, the inner disk disappeared due to PEWs, and the planets stopped accreting atmospheres. 
Figure~\ref{fig:mcore_menv2}(b) summarizes five runs of simulations for PL2, which shows that the final atmospheric mass fractions were less than $\sim$10\%. Thus, we also found that super-Earths without massive atmospheres could form without pebble accretion in disks that underwent rapid disk clearing via PEWs.

Figure~\ref{fig:t_a2}(c) shows an outcome of simulation EM3 in which the initial total solid mass was increased by a factor of two relative to the PL2 simulations. Compared to the simulation shown in Figure~\ref{fig:t_a2}(b), the solid surface density was higher, and thus the isolation mass was larger ($\simeq 5 \,M_\oplus$). The isolation mass was larger than the critical core mass; therefore the cores started accreting atmospheres during the early phase ($t<0.1 {\rm \,Myr}$). As the cores could accrete atmospheric gas for about 1\,Myr, the super-Earth cores possessed more massive atmospheres than those in the PL2 simulations.
As shown in Figure~\ref{fig:mcore_menv2}(c), the planets had up to about 30wt\% atmospheres, and the accreted masses were larger than those shown in Figure~\ref{fig:mcore_menv2}(b) for PL2. However, it is worth noting that the atmospheric mass fractions were smaller than those found from the simulations of disks that did not undergo mass loss due to PEWs (EM1), even though the core masses were larger. Therefore, we can conclude that rapid disk clearing by PEW plays a role in limiting, or indeed inhibiting, the accretion of massive H$_2$/He atmospheres from the remnant disk.

\begin{figure}[ht!]
\epsscale{1.0}
\plotone{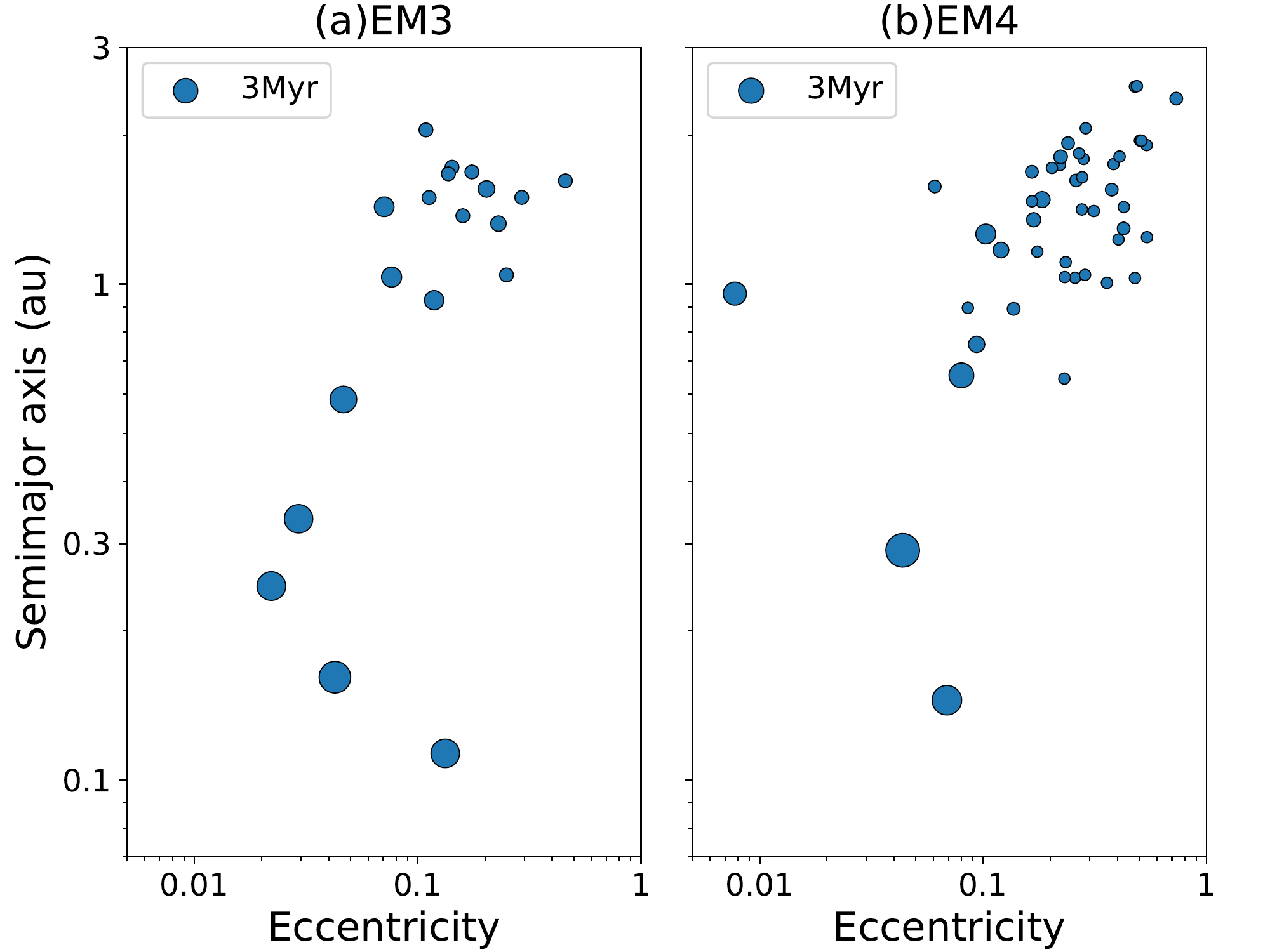}
\caption{Snapshots of systems at $t = 3 {\rm \,Myr}$ for EM3 and EM4. The size of circles are proportional to $M^{1/3}$.}
\label{fig:a_e}
\end{figure}

We also performed simulations in which the initial mass of the embryo was decreased by a factor of three (EM4) relative to the PL3 simulations, although we maintained the initial solid surface density. That is, the initial embryo mass was set to $M_{\rm ini} = 0.03 \,M_\oplus$, and the number of initial particles was increased by a factor of three. Figure~\ref{fig:t_a2}(d) shows the temporal evolution of a typical run. As mentioned above, the isolation mass was larger than the critical core mass. Thus, the planets accreted atmospheres during the early stage. The big difference from EM3 was that planets possessed less massive H$_2$/He atmospheres, which can clearly be seen when comparing Figures~\ref{fig:mcore_menv2}(c) and (d). As seen in Figure~\ref{fig:t_a2}(d), this is because the planets experienced more impact erosion events after the gas disk dissipated at $t \gtrsim 1.5 {\rm \,Myr}$.
As a consequence of viscous stirring and energy equipartition between cores with $M \simeq 1 \,M_\oplus$ and small embryos with $M \simeq 0.03 \,M_\oplus$, the eccentricities of the small embryos were excited. Such small embryos in the inner region ($r \simeq 0.3 {\rm \,au}$) were accreted by the cores before $t = 1 {\rm \,Myr}$; however, high-eccentricity ($e \gtrsim 0.3$) embryos remained in the outer regions ($r \gtrsim 1 {\rm \,au}$), even after the disk gas dissipated. Figure~\ref{fig:a_e} compares a snapshot of the system at $t = 3 {\rm \,Myr}$ with the same epoch from EM3. The existence of high-eccentricity embryos in outer orbits ($r \gtrsim 1 {\rm \,au}$) can be clearly seen in the EM4 simulations. These high-eccentricity embryos collided with the inner super-Earths, leading to impact erosion of the accreted atmospheres. This process is discussed in detail in Section~\ref{sec:discussion2}.

\section{Discussion}\label{sec:discussion}
\subsection{Condition for avoiding runaway gas accretion}\label{sec:discussion1}

The condition for accreting a massive H$_2$/He atmosphere is that once a core grows large enough ($M_{\rm core} \gtrsim 3 \,M_\oplus$), some disk gas must be present. In this section, we discuss in more details the results obtained in Section~\ref{sec:results}.

We will first discuss the core masses in comparison with the critical core mass. As pebbles accrete onto a core, the critical core mass remains large ($\gtrsim 10 \,M_\oplus$) due to pebble heating. Runaway gas accretion is usually avoided during this phase unless the core mass is significantly large ($M_{\rm core} > 10 \,M_\oplus$). This has been shown by \citetalias{2020ApJ...892..124O}. The critical core mass with pebble heating was given by Eq.~(\ref{eq:mcrit}).

\begin{figure}[ht!]
\plotone{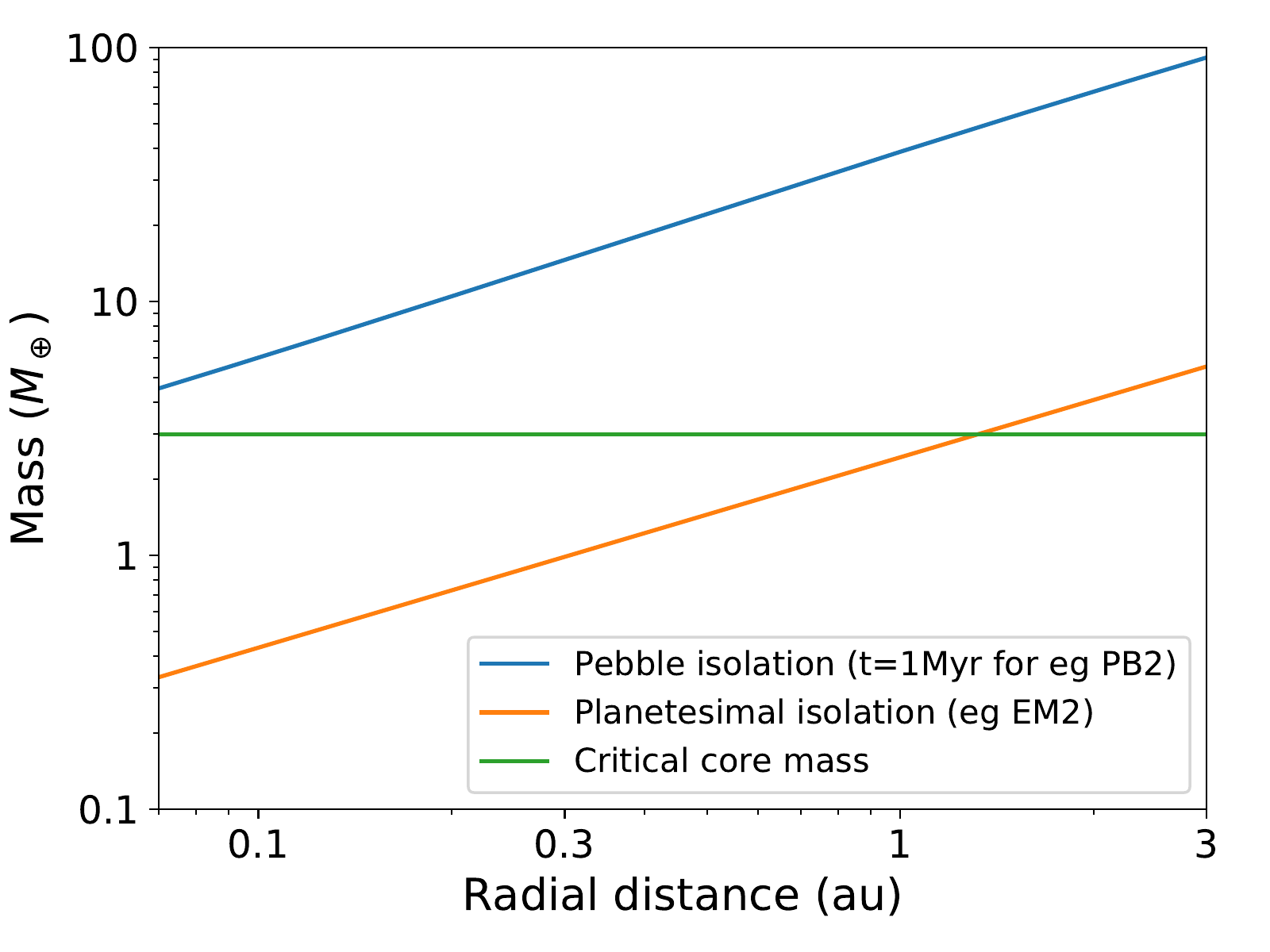}
\caption{The pebble isolation masses calculated from Eq.~(\ref{eq:miso_pb}) in our standard disk under the effects of PEWs at $t = 1 {\rm \,Myr}$, in which the disk temperature is primarily determined by the stellar irradiation in the late stage of disk evolution (see fig.\,1 of \citealt{2016A&A...596A..74S}). The planetesimal isolation masses were calculated using Eq.~(\ref{eq:miso_pl2}), in which $\Sigma_{\rm d} = 77 {\rm \,g \,cm^{-2}} (r/1 {\rm \,au})^{-3/2}$ was assumed. The effective critical core mass ($= 3 \,M_\oplus$) is also shown (see Section~\ref{sec:atm_acc}).}
\label{fig:miso}
\end{figure}

Planetary cores cease solid accretion once they reach the isolation mass (either $M_{\rm iso,pb}$ or $M_{\rm iso,em}$), as seen in previous studies \citep[\citetalias{2020ApJ...892..124O},][]{2020arXiv200604121B}. If the core mass is sufficiently small ($\lesssim 2 \,M_\oplus$), the core does not accrete a massive atmosphere. Figure~\ref{fig:miso} compares the isolation masses and the critical core masses. As stated in Section~\ref{sec:atm_acc}, the critical core mass was effectively larger than $3 \,M_\oplus$, and the pebble isolation mass was usually larger than the critical core mass without pebble heating. On the other hand, the isolation mass for planetesimal accretion was smaller than $2 \,M_\oplus$ at $r \lesssim 1 {\rm \,au}$, assuming $\Sigma_{\rm d} = 77 {\rm \,g\,cm^{-2}} (r/1 {\rm \,au})^{-3/2}$ (Eq.~\ref{eq:miso_pl}). Thus, the cores did not exceed the critical core mass in this case (see Figure~\ref{fig:miso} for the isolation mass).

After cores reached the isolation mass, they can grow via giant impacts between planets. After this giant-impact phase, the final core masses were effectively determined. When pebble accretion was considered, the pebble isolation mass was usually larger than $\simeq 3 \,M_\oplus$, and the core started accreting a massive atmosphere before the gas-dissipation phase. The value of $\Delta a/ R_{\rm H}$, where $\Delta a$ denotes the orbital separation, became smaller while growing further by atmospheric accretion and pebble accretion (and hence increasing $R_{\rm H}$), which led to giant impacts in the gas disk (e.g., PB1 and PB2). 
On the other hand, when planets grow solely by accreting embryos, a late orbital instability was triggered via dissipation of the disk gas after the eccentricity damping effect had weakened. In the EM1 and PL2 simulations, the typical gas surface density for triggering the late orbital instability was $\Sigma_{\rm g,GI} \sim 1 {\rm \,g \,cm^{-2}}$. If the core masses were still small after the giant impacts, the planets did not eventually experience any runaway gas accretion. To form super-Earths, the core mass should be $\gtrsim 3 \,M_\oplus$, which was larger than the critical core mass. Therefore, the core eventually exceeded the critical core mass.

Next, we will discuss the disk accretion rate when the core mass exceeded the critical core mass. Even if the core mass exceeded the critical core mass, the core could avoid accreting a massive H$_2$/He atmosphere if there was only small amount of gas accretion onto the core. In the following, we will consider two cases.
First, we considered a case in which the core exceeded the critical core mass in the disk-dissipation phase ($t \gtrsim 1 {\rm \,Myr}$) (e.g., PB1--PB2 and EM1--EM2). In a disk not subjected the effects of PEWs, the disk accretion rate was as high as $\dot{M}_{\rm disk} \simeq 10^{-4} \,M_\oplus \,{\rm yr^{-1}}$, even at $t \gtrsim 3 {\rm \,Myr}$ and $\Sigma_{\rm g} = \Sigma_{\rm g,GI} (\sim 1 {\rm \,g\,cm^{-2}})$ (see Figure~\ref{fig:sgm_evol} for the temporal evolution of the disk accretion rate). Thus, the core accreted a massive atmosphere in PB1 and EM1\footnote{Note also that even when the disk dissipated slowly in the absence of PEW effects, the effective disk accretion rate was small if the wind-driven accretion did not contribute to atmospheric accretion \citep{2018ApJ...867..127O}.}. However, in disks that underwent mass loss due to PEWs, the disk-clearing timescale after the disk started to dissipate was quite short ($\sim 0.1 {\rm \,Myr}$), and the disk accretion rate soon became very small at $t \gtrsim 1.5 {\rm \,Myr}$ (see the bottom panel of Figure~\ref{fig:sgm_evol}). In this scenario, the cores did not accrete massive atmospheres (PB2 and EM2). Thus, the short disk-dissipation timescale due to PEWs was important for inhibiting the accretion of a massive atmosphere.
Second, we considered a case in which the core exceeded the critical core mass well before the disk-dissipation phase (e.g., PB3 and EM3). If the core exceeded the critical core mass during the early phase when the disk accretion rate was high ($\gtrsim 10^{-4} \,M_\oplus \,{\rm yr^{-1}}$), it started accreting a massive atmosphere. Nevertheless, it is worth noting that the amount of accreted gas was less than that in disks not subjected to mass loss due to PEWs (e.g., PB3 and EM3), because the duration of atmospheric accretion was shorter than in PB1 and EM1. Therefore, in both cases, the disk dissipation due to PEWs played an important role in decreasing the amount of accreted gas.

\subsection{Impact erosion by eccentric embryos}\label{sec:discussion2}

We found that impact erosion after gas dissipation was more effective in the EM4 simulation (Figure~\ref{fig:t_a2}(d)), in which the initial mass of the embryos was smaller than in EM3. Here we will discuss the reason for this. 
First, we should consider that embryos in eccentric orbits were present after the disk dissipated. The excitation and damping of the average eccentricity ($e_2$) of a population of embryos with $M=M_2$ that interacted with a population of embryos with $M=M_1$ and an average eccentricity of $e_1$ can be expressed as \citep{1993prpl.conf.1089O}
\begin{eqnarray}\label{eq:dedt}
\frac{de_2^2}{dt}  &\simeq& \left(\frac{M_1}{M_1 + M_2}\right)^2 
\left( \frac{e_1^2 + e_2^2}{t_{\rm g,21}} + \frac{M_1 e_1^2 - M_2 e_2^2 }{M_1 t_{\rm g,21}} \right) \nonumber \\
&&+  \frac{e_2^2}{2 t_{\rm g,22}},\\
&\sim& \frac{2 e_1^2 + e_2^2}{t_{\rm g,21}} + \frac{e_2^2}{2t_{\rm g,22}},
\end{eqnarray}
where the first and second terms on the right-hand side indicate viscous stirring and dynamical friction of a population with $M_1$, respectively. The third term indicates stirring from embryos in the same population with $M_2$. For the last deformation, $M_1 \gg M_2$ was assumed. The timescale of the change in the random velocity due to encounters with the population of $M_1$ can be given by
\begin{eqnarray}
t_{\rm g,21}  \simeq \frac{M_1}{\pi \Sigma_1 \ln \Lambda} \left[ \frac{v^2}{G (M_1 + M_2)} \right]^2 \Omega_{\rm K}^{-1},
\end{eqnarray}
where $\ln \Lambda$ is the Coulomb logarithm. Given that $t_{\rm g,21} \propto (\Sigma_1 M_1)^{-1}$ and $t_{\rm g,22} \propto (\Sigma_2 M_2)^{-1}$, the eccentricities were determined mainly by the population of embryos with $M_1$ for $\Sigma_1 M_1 > \Sigma_2 M_2$. 
The eccentricities could also be damped by gravitational interactions with the disk gas. This damping timescale is given by
\begin{eqnarray}\label{eq:tdamp}
t_{\rm damp} \simeq \left(\frac{M}{M_*}\right)^{-1} \left(\frac{\Sigma_{\rm g} r^2}{M_*}\right)^{-1} \left(\frac{c_{\rm s}}{v_{\rm K}}\right)^4 \Omega_{\rm K}^{-1}.
\end{eqnarray}
which indicates that eccentricity damping is weaker for lower-mass embryos, and therefore, smaller embryos could maintain high eccentricities. 

If such high-eccentricity embryos were not accreted by the cores for long periods, they could exist after the disk dissipated. The core growth timescale is given by Eq.~(\ref{eq:tacc_pl}), which shows that the core accretion timescale is proportional to $r^{3}$ assuming $\Sigma_{\rm d} \propto r^{-3/2}$. Hence, small embryos could remain in the outer orbits after gas dispersal. Note also that if the eccentricities were damped by the gas disk, the embryos could have been more efficiently accreted by the cores due to gravitational focusing.

To summarize, the eccentricities of the embryos were determined by viscous stirring, dynamical friction, and gas damping. Smaller embryos could maintain high eccentricities because the gas damping effect was less efficient. Such embryos could survive even after gas dispersal because core accretion was less efficient in the outer regions. As a result, the inner super-Earths experienced impact erosion more frequently when smaller embryos were adopted as the initial condition. 

In this paper, the outer boundary of the initial embryo distribution was set to be 1\,au. If simulations start with embryos at larger distances ($r > 1 {\rm \,au}$), there can be an additional source of impactors that cause the impact erosion. Collisions of icy embryos that form in the outer region is also interesting in terms of water delivery to inner super-Earths. Although \textit{N}-body simulations that consider small embryos in the outer region, which avoid efficient eccentricity damping, are computationally demanding, this should be investigated in a separate study.

\begin{figure}[ht!]
\epsscale{1.1}
\plotone{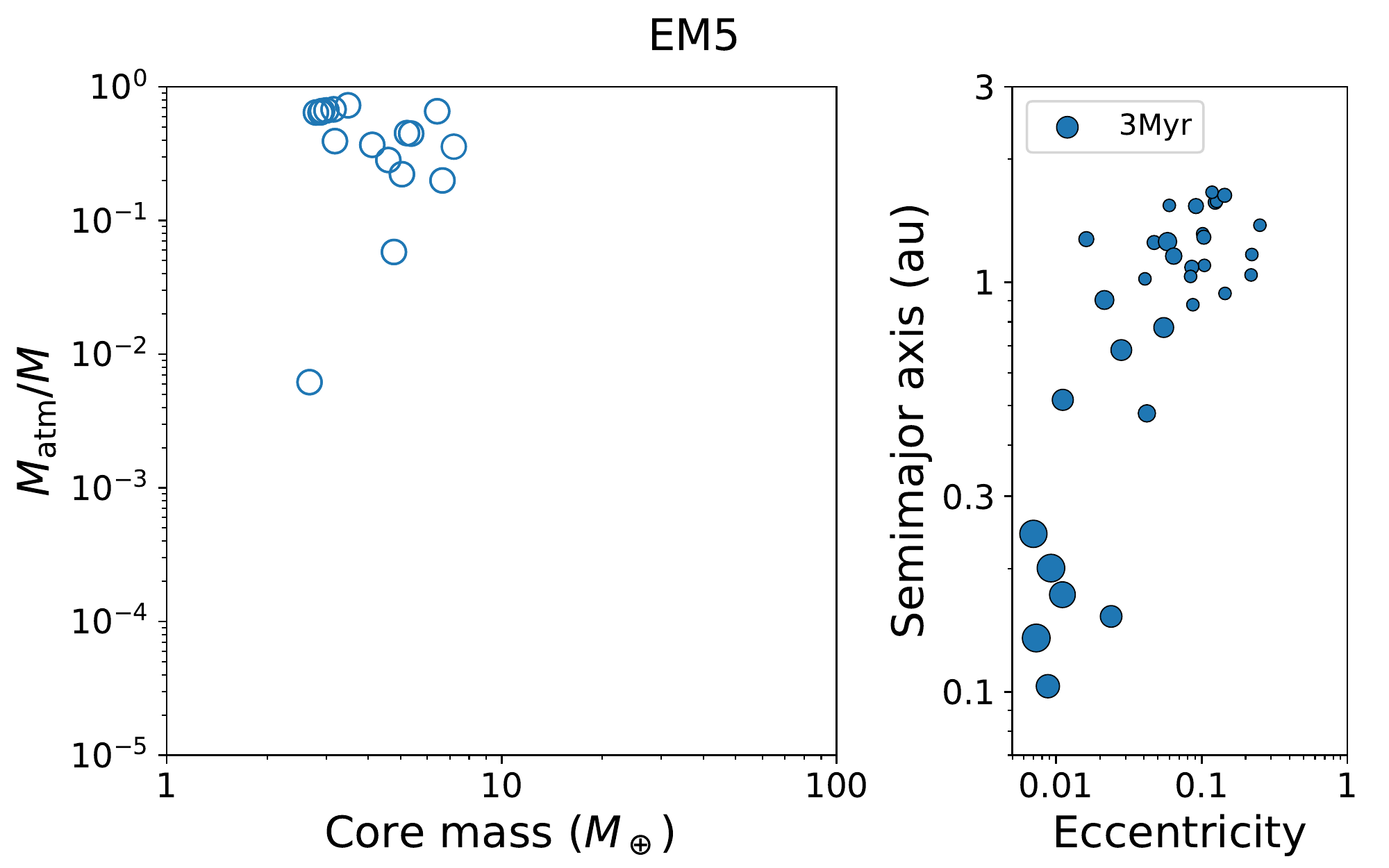}
\caption{{\it Left.} Atmospheric mass fractions at the end of five simulation runs for EM5. {\it Right.} A snapshot at $t = 3 {\rm \,Myr}$ of a typical run for EM5. The size of the circles are proportional to $M^{1/3}$.}
\label{fig:mcore_menv_run471}
\end{figure}

Next, we will discuss whether impact erosion due to high-eccentricity embryos is also effective in disks without the mass loss due to PEWs. We performed additional simulations (EM5) in which $M_{\rm ini}$ was decreased by a factor of three relative to model EM1. The left panel of Figure~\ref{fig:mcore_menv_run471} shows a summary of five runs of simulations. Compared with Figure~\ref{fig:mcore_menv2}(a) for EM1, we found that the atmospheric fractions were slightly smaller due to impact erosion. However, this difference was very small, and the super-Earths possessed massive atmospheres at the end of the simulations. The number of high-eccentricity embryos was smaller in the disk-dissipation phase, and their eccentricities were also smaller ($e \lesssim 0.3$), as shown in the right panel of Figure~\ref{fig:mcore_menv_run471}. This is because the disk lifetime was longer in the simulations without mass loss due to PEWs, and the eccentricities of the embryos were damped by the remnant disk. Therefore, rapid disk clearing due to the PEW is required to decrease the amount of accreted atmospheres by impact erosion of high-eccentricity embryos.

It could be possible that impact erosion could remove a large amount of accreted gas for cases with pebble accretion, given that high-eccentricity embryos remained after disk dispersal. Note that, as seen in the EM4 simulations, a certain number of high-eccentricity particles were needed to significantly reduce the atmosphere masses. We leave investigation of this mechanism for future work.

In this study, we considered impact erosion during giant impacts. That is, the difference between the target mass and impactor mass was up to a factor of about 10. It may be possible that the mass loss due to smaller planetesimal impacts plays a role in atmospheric loss \citep[e.g.,][]{2015Icar..247...81S}. It remains to be investigated whether a large number of high-eccentricity planetesimals exist after gas dispersal.

\subsection{Model caveats}\label{sec:discussion3}

In this paper, we used the disk evolution model developed by \citet{2016A&A...596A..74S} and \citet{2020MNRAS.492.3849K}. Although this model are based on (magneto)hydrodynamical simulations, it has some uncertainties. For example, the parameter for the turbulent viscosity, $\arp$, is not well constrained. To compare with \citetalias{2020ApJ...892..124O}, we used the same value of $\arp$. We also performed additional simulations using smaller value of $\arp$ in Appendix~\ref{sec:app}. As a result, our conclusions did not change. That is, due to rapid disk clearing by PEWs, super-Earth cores can avoid accreting massive atmospheres.

We also briefly considered a different case. If the effects of MDWs, including wind-driven accretion, and PEWs are weaker, it is expected that the disk lifetime will become longer \citep{2020MNRAS.492.3849K}. Although we did not perform simulations for such a case, we can comment on whether longer disk lifetimes could changes our conclusions.
If the disk lifetime was longer in simulations with pebble accretion (e.g., PB2), it is likely that the amount of accreted gas increased. This is because the time span of atmospheric gas accretion is longer than that for PB2. Note, however, that as the inner disk was still cleared out very rapidly ($\sim 0.1 {\rm \,Myr}$), the time span of atmospheric accretion was shorter than that in disks not subjected to the effects of PEWs (e.g., PB1).
If the disk lifetime was longer in simulations without pebble accretion (e.g., EM2), it is expected that the amount of accreted gas would not change greatly. As we have seen in the PB2 simulations, the core masses usually did not exceed the critical core mass in the disk. Instead, a core started runaway gas accretion only when the disk dissipated to some extent ($\Sigma_{\rm g} \sim 1 {\rm \,g \,cm^{-2}}$), which then subsequently grew via giant impacts. As the onset of runaway gas accretion was also delayed in disks with longer lifetimes, we expect that the amount of accreted gas does not change greatly.
These cases will be explored further in future work.

\section{Conclusions}\label{sec:conc}

We examined the effects of rapid disk clearing due to PEWs on super-Earth atmospheres. Our main findings can be summarized as follows.
\begin{enumerate}
\item Due to rapid disk clearing by PEWs, the duration of gas accretion onto the cores became shorter, which prevented the cores from accreting massive atmospheres. In previous disk models, in which mass loss due to PEWs was not considered, the disk gas gradually dissipated over a timescale of a few Myr. Hence, cores that reached the critical core mass had enough time to accrete massive atmospheres. On the other hand, in the new disk model in which the mass loss by PEWs was considered, the disk rapidly cleared out during the disk-dissipation phase. As a result, the cores could only accrete gas for a short time. Such rapid disk dissipation has also been observed \citep[][]{2014prpl.conf..475A}.

\item When the cores grew solely by planetesimal accretion, they grew rapidly in the disk. Nevertheless, the cores could avoid accreting massive atmospheres. If the cores grew by pebble accretion, they stopped solid accretion when they reached the pebble isolation mass ($\gtrsim 10 \,M_\oplus$), which was larger than the critical core mass because there was no heating in their atmospheres \citepalias{2020ApJ...892..124O} that could trigger the onset of runaway gas accretion before the disk-dissipation phase. On the other hand, if the cores grew solely via planetesimal accretion, the isolation mass ($\lesssim 2 \,M_\oplus$) was smaller than the critical core mass in some cases. Thus, the cores exceeded the critical core mass only when they experienced giant impacts during the disk-dissipation phase. Therefore, even if the cores grew rapidly by planetesimal accretion, the onset of runaway gas accretion was delayed. Together with the rapid disk clearing by the PEWs, the cores avoided accreting massive atmospheres. Note that even if the onset of runaway accretion was delayed, the cores could accrete massive atmospheres from the remnant disk if the rapid disk clearing by PEWs was not considered.

\item Due to the rapid clearing of the disks, there exist some high-eccentricity embryos after disk dissipation, which led to efficient impact erosion of the accreted atmospheres. In the work of \citetalias{2020ApJ...892..124O}, it was shown that giant-impact events between super-Earths are not frequent; thus the accreted massive atmospheres could not be efficiently reduced by impact erosion. In this study, however, we found that the accreted atmospheres could be lost efficiently by impact erosion if small, high-eccentricity embryos were present after the disk dissipated. Such high-eccentricity embryos collided with the super-Earths in inner orbits, which stripped their accreted atmospheres. This mass-loss mechanism may be more important than atmospheric photoevaporation because atmospheric losses via photoevaporation is efficient only for planets in very close-in orbits ($P \lesssim 10 {\rm \,day}$). Note that for disks that did not experience rapid clearing by PEWs, the high-eccentricity embryos were less likely to remain after disk dissipation. This is because the remnant disk gas damped the eccentricities of the embryos, and such embryos with small eccentricities were efficiently accreted by the cores due to strong gravitational focusing. If the disk was dissipated rapidly by PEWs, high-eccentricity embryos could remain in the outer regions ($\simeq 1 {\rm \,au}$).
\end{enumerate}

We demonstrated that super-Earths were prevented from accreting massive atmospheres irrespective of the accretion mode. It would be very interesting if we can determine whether super-Earths underwent pebble accretion in the past from observation. However, it is difficult to do so, because planets with typical masses of super-Earths can form in both cases with and without pebble accretion. Comparing Figures~\ref{fig:mcore_menv} and \ref{fig:mcore_menv2}, cores with less than about $10 \,M_\oplus$ have lower-mass atmospheres in cases with pebble accretion. This feature may disappears when we consider atmospheric accretion onto cores with $M < M_{\rm crit}$ (see Section~\ref{sec:pebble_PEW}). This will be a subject of future studies.

We investigated formation of super-Earths in inner orbits inside 1~au. It would be interesting to investigate formation of giant planets at larger distances. Due to the effects of MDWs (mainly wind-driven accretion), the surface density slope of gas disk is shallower than that for the minimum-mass solar nebula. Thus, the inward migration of giant planet cores can be slowed down even in the outer region, which may help to form giant planets. However, it would still be difficult to form giant planet cores within the disk lifetime \citep[e.g.][]{2019A&A...631A..70J}. A sufficiently high pebble flux \citep{2019A&A...623A..88B} and/or enhanced planet-planet collisions in a warmer disk \citep{2020MNRAS.496.3314W} would be required for the growth of large cores in wide orbits.

\acknowledgments

We thank two referees for valuable comments. We also thank Hiroshi Kobayashi for useful discussions.
This work was supported by JSPS KAKENHI Grant Numbers 17H01105, 18K13608, 19H05087, and 20K14542. The numerical computations were, in part, carried out on the PC cluster at the Center for Computational Astrophysics of the National Astronomical Observatory of Japan.

\appendix
\section{Case for lower $\arp$}\label{sec:app}

In the work of \citet{2016A&A...596A..74S}, disk evolution was simulated for two values of $\arp$. In this study, we adopted disk evolution for a higher value of $\arp (= 8 \times 10^{-3})$ as a standard value. In such a disk, the surface density slope is almost flat in the inner region inside 1\,au, and type-I migration can be significantly suppressed. Many observed orbital properties (e.g., the period--ratio distribution) can be reproduced in this case \citep[][]{2018A&A...615A..63O}. On the other hand, when a lower value of $\arp (= 8 \times 10^{-5})$ is used, the surface density slope can be positive in the close-in region, leading to outward migration of embryos and planetesimals \citep[e.g.,][]{2018A&A...612L...5O}. Here, we briefly mention the model in which the lower value of $\arp$ was used, where the gas was carved out from the inside.

\begin{figure}[ht!]
\epsscale{0.6}
\plotone{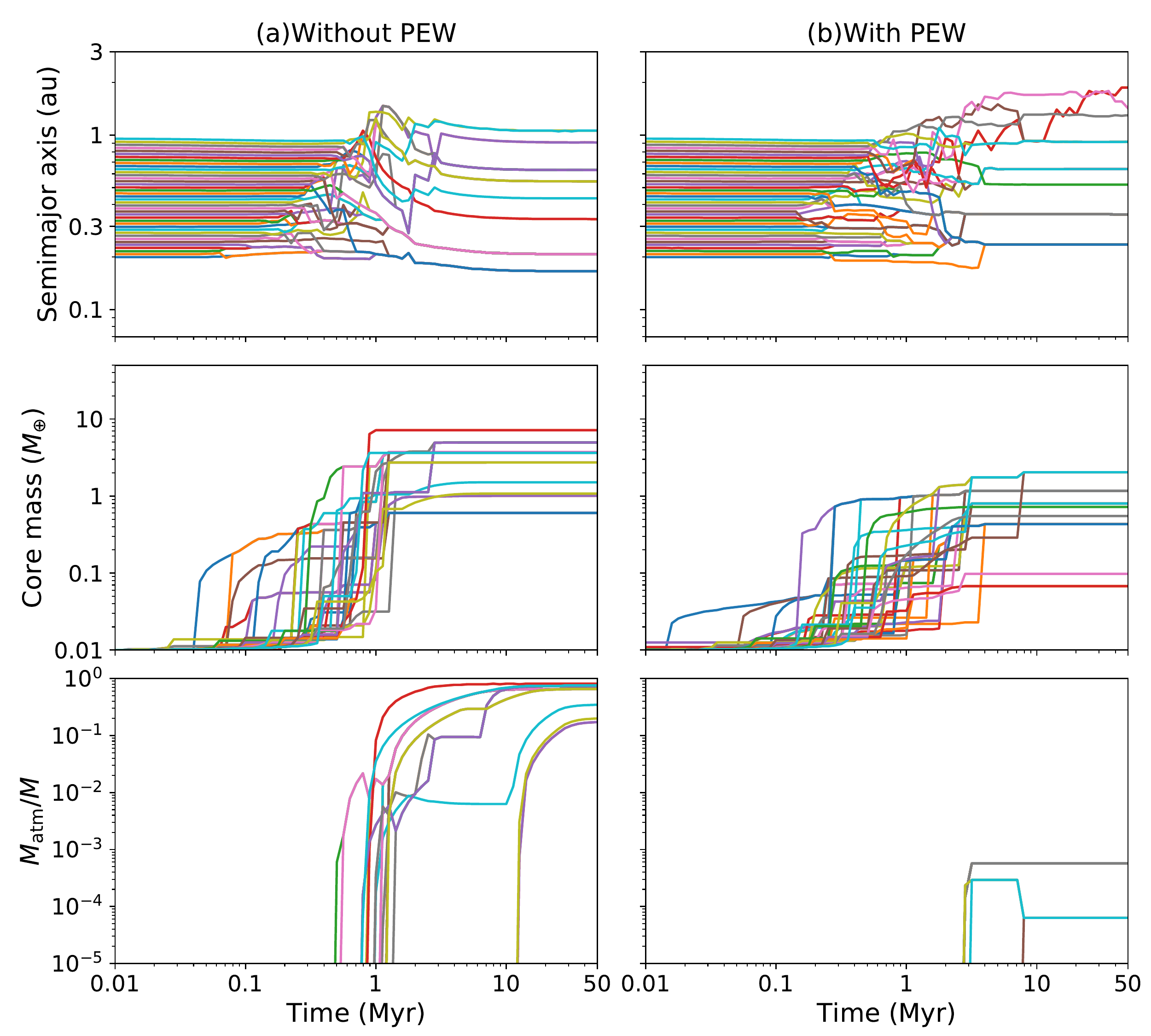}
\caption{Same as Figure~\ref{fig:t_a1}, but for inactive disks.}
\label{fig:t_a3}
\end{figure}

\begin{figure}[ht!]
\epsscale{0.6}
\plotone{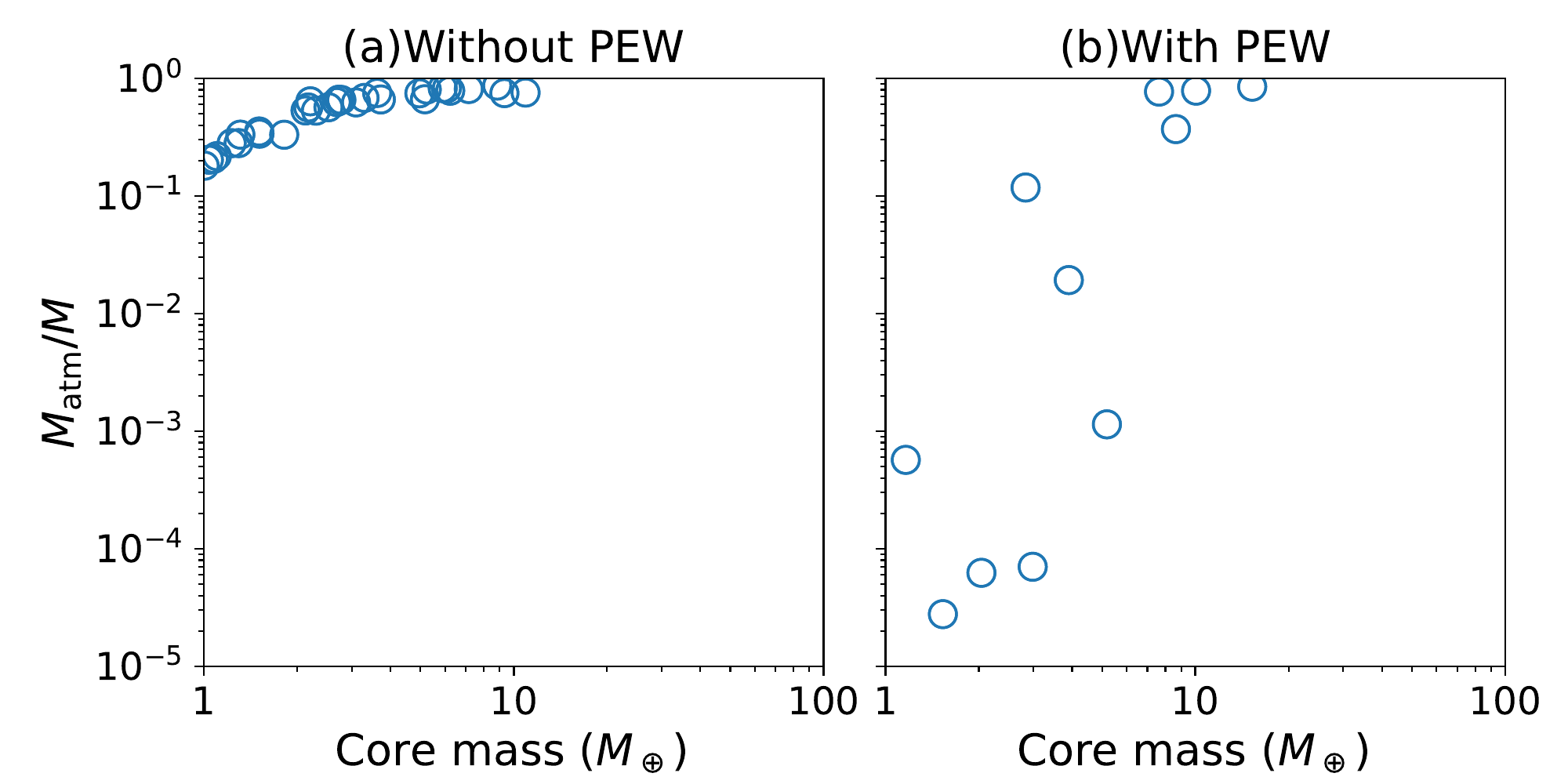}
\caption{Same as Figure~\ref{fig:mcore_menv}, but for inactive disks.}
\label{fig:mcore_menv3}
\end{figure}

We confirm that our findings hold true for this different disk evolution (see figure\,2 of \citet{2020MNRAS.492.3849K} for the corresponding disk evolution). Figure~\ref{fig:t_a3} shows typical results for disks without and with mass loss due to PEWs. Note that the pebble flux was set to a lower value of $\dot{M}_{\rm pb} = 5 \times 10^{-6} M_\oplus {\rm \,yr^{-1}}$, because pebble accretion is more efficient for smaller pebble scale heights.
After the planetary cores grew to the pebble isolation mass, they accreted massive H$_2$/He atmospheres in disks without PEW effects (Figure~\ref{fig:t_a3}(a)). When the disk evolution includes PEWs, the time span of gas accretion was shortened, and super-Earths could form without accreting massive atmospheres (Figure~\ref{fig:t_a3}(b)). We also found that the migration was suppressed compared to the previous study \citep{2018A&A...615A..63O}, who did not consider pebble accretion. 
Figure~\ref{fig:mcore_menv3} summarizes five runs of simulations. We found that the super-Earths possessed a large amount of atmospheric gas in disks that did not contain PEWs (Figure~\ref{fig:mcore_menv3}(a)). On the other hand, if the inner disk was rapidly cleared out by the PEWs, the amount of accreted gas decreased (Figure~\ref{fig:mcore_menv3}(b)). This trend is the same as seen in Section~\ref{sec:pebble}.

\bibliography{reference}
\bibliographystyle{aasjournal}



\end{document}